\documentclass[english]{article}
\usepackage[T1]{fontenc}
\usepackage[latin9]{inputenc}
\usepackage{amsmath}
\usepackage{amsthm}
\usepackage{amssymb}
\usepackage{graphicx}

\makeatletter
\theoremstyle{plain}
\newtheorem{thm}{\protect\theoremname}

\usepackage{babel}
\providecommand{\theoremname}{Theorem}

\usepackage{babel}
\providecommand{\theoremname}{Theorem}

\usepackage{babel}
\providecommand{\theoremname}{Theorem}

\makeatother

\usepackage{babel}
\providecommand{\theoremname}{Theorem}

\begin{document}
\title{Quasi-Objective Coherent Structure Diagnostics \\
 from Single Trajectories}
\author{George Haller\thanks{Email:georgehaller@ethz.ch}, Nikolas Aksamit
and Alex P. Encinas-Bartos\\
 Institute for Mechanical Systems\\
 ETH Zürich, 8092 Zürich, Switzerland\\
 }
\maketitle
\begin{abstract}
We derive measures of local material stretching and rotation that
are computable from individual trajectories without reliance on other
trajectories or on an underlying velocity field. Both measures are
quasi-objective: they approximate objective (i.e., observer-independent)
coherence diagnostics in frames satisfying a certain condition. This
condition requires the trajectory accelerations to dominate the angular
acceleration induced by the spatial mean vorticity. We illustrate
on examples how quasi-objective coherence diagnostics highlight elliptic
and hyperbolic Lagrangian coherent structures even from very sparse
trajectory data. 
\end{abstract}
\begin{quote}
\textbf{~~~~~~~~~~~~~~~~~~~~~~~~~~~~~~~~~~~~~~~~~~~~~Summary}

\textbf{A vast amount of discrete tracer trajectory data is available
for fluid flows in the laboratory and in nature. All features of tracer
trajectories, however, including their shape, velocities and looping,
depend on the reference frame of the observer even though the coherent
flow structures one often tries to infer from tracers are observer
indifferent. Here we develop quasi-objective coherent structure diagnostics
that can be computed from individual single tracer trajectories without
reliance on other trajectories. A quasi-objective diagnostic approximates
an objective diagnostics in all frames of reference that satisfy certain
conditions we derive. We show how the single-trajectory quasi-objective
scalar fields highlight material coherent structures even from sparse
trajectory data sets on which multi-trajectory objective diagnostics
tend to perform poorly.} 
\end{quote}

\section{Introduction}

We seek objective (i.e. observer-indifferent) diagnostics for material
stretching and rotation that can be evaluated on individual trajectories
without reliance on other trajectories or on the underlying velocity
field. Applications calling for such diagnostics include the analysis
of observational drifter data, balloon data and particle tracks from
particle tracking velocimetry (PTV).

All elementary features of particle paths are non-objective, i.e.,
depend on the observer. For instance, in a frame traveling with any
given particle, the particle becomes just a fixed point. All features
of the trajectory, such as velocity, acceleration, looping number,
curvature and trajectory length, therefore, vanish in this frame.
In contrast, Lagrangian coherent structures (LCS), the persistent
features of the material deformation field of the fluid as a moving
continuum, are objective by definition \cite{haller15}. Accordingly,
any self-consistent identification of LCS should be carried out using
objective quantities. Indeed, objectivity as a minimal requirement
for flow-feature identification was already identified in the 1970's
\cite{drouot76a,drouot76b,astarita79,lugt79} and a number of recent
approaches conform to this requirement (see \cite{haller05,haller15,peacock15,kirwan16,gunther18}
for reviews).

Objective geometric quantities used in LCS identification are, however,
based on the gradient of the flow map, whose computation requires
a large number of trajectories to be released from an initial grid.
Examples of such quantities include the finite-time Lyapunov exponent
(or FTLE, \cite{haller01}) for hyperbolic LCS and the polar rotation
angle (or PRA, \cite{farazmand16}) for elliptic LCS in two dimensions.
The computation of these diagnostics is only feasible if the velocity
field is known over, or can at least can be interpolated onto, a regular
grid over a full flow domain \cite{haller15}. For the special case
of closed (or elliptic) LCSs, alternative probabilistic methods have
been developed that seek regions enclosed by such LCSs as coherent
sets \cite{froyland10,froyland13}. These objective methods do not
technically require differentiation of the flow map, but their implementations
utilize box-counting methods that only converge for large numbers
of densely spaced trajectories. More recent clustering methods for
coherent set detection are, in principle, applicable to sparse trajectory
data, but still require a reasonably high number of evenly placed
trajectories to give meaningful results \cite{hadjighasem16,froyland15,Schlueter-Kuck17}.
None of these methods offer a coherence indicator that is computable
on a single trajectory. Rather, the coherence measures assigned to
individual trajectories by probabilistic and statistical methods will
change if the position and number of the other trajectories used in
the algorithm change--even though the underlying fluid flow and its
coherent structures remain the same.

Notable exceptions to this trend are two objective indicators of the
complexity of individual trajectories: the correlation dimension and
the ergodicity defect proposed in \cite{rypina11} for LCS extraction.
While these objective diagnostics can be computed from single trajectories,
their connection to material attributes of fluid trajectory motion,
such as material stretching and rotation, are unknown. Absolute dispersion
\cite{provenzale99}, trajectory length \cite{mancho13} and maximal
trajectory extent \cite{mundel14} are also attractively simple single-trajectory
diagnostics, but they are not objective and their relationship to
material stretching and rotation is also unknown. In contrast, the
Lagrangian-averaged vorticity deviation (LAVD) is an objective single-trajectory
diagnostic \cite{haller16b} but requires the knowledge of the full
velocity field for the purposes of computing the vorticity along trajectories.
A further non-objective, single-particle diagnostics for LCS identification
is the Lagrangian spin parameter \cite{veneziani04}, which relies
on the calibration of an additional stochastic model for the velocity
field. Yet another approach adapts the wavelet ridge analysis method
of \cite{delprat92} to Lagrangian trajectory analysis \cite{lilly06,lilly20}.
While physically insightful, this technique is non-objective either,
as it fits a time-varying ellipse model to identify signatures of
coherent eddies based on the looping characteristics of trajectories.

While all identifiable features of individual trajectories are non-objective,
trajectory data is often the only available information about fluid
flows. It is therefore essential to assess what trajectory-based quantities
(if any) can still be tied mathematically, at least in qualifying
frames, to established and objective LCS diagnostics. Answering this
question would enable, for instance, a mathematically justifiable
and physically self-consistent extraction of Lagrangian jets, fronts
and eddies from NOAA's Global Drifter Program, which comprises more
than 20,000 trajectories \cite{lumpkin07}. Such an extraction should
then lead to a systematic assessment of the performance of promising
nonobjective trajectory-based methods, such as \cite{lilly20}, in
avoiding false positives and negatives in elliptic LCS detection.

Here we address this challenge by deriving objective coherence diagnostics
for material stretching and rotation that are computable solely from
single-particle trajectory data. Each diagnostic approximates a related
objective, material coherence measure in frames that satisfy certain
computable conditions. We refer to the approximate coherence diagnostics
developed here as quasi-objective\emph{ }under those conditions. Physically,
the quasi-objectivity condition we obtain requires the trajectory
accelerations to dominate the angular acceleration induced by the
spatial mean vorticity in the given frame. We illustrate the power
of quasi-objective stretching and rotation measures on two- and three-dimensional
examples.

\section{Objective Lagrangian stretching and rotation measures\label{sec:Objective-Lagrangian-measures}}

Consider particle motions\textbf{ }generated by a differential equation
\begin{equation}
\dot{\mathbf{x}}=\mathbf{v}(\mathbf{x},t),\label{eq:velocity field}
\end{equation}
with a continuously differentiable velocity field $\mathbf{v}$ defined
for times $t\in[t_{0},t_{N}]$ on a spatial domain $U\subset\mathbb{R}^{n}$
of dimension $n\geq1$. Let \textbf{$\mathbf{x}(t;t_{0},\mathbf{x}_{0})$}
be a trajectory with initial condition $\mathbf{x}(t_{0};t_{0},\mathbf{x}_{0})=\mathbf{x}_{0}$.
The flow map induced by trajectories is $\mathbf{F}_{t_{0}}^{t}\colon\mathbf{x}_{0}\to\mathbf{x}(t;t_{0},\mathbf{x}_{0})$,
mapping initial conditions at time $t_{0}$ to their current positions
at time $t$. When $t_{0}$ and $\mathbf{x}_{0}$ are fixed and hence
do not need to be carried in our arguments, we will also use the shorthand
notation $\boldsymbol{x}(t):=\mathbf{x}(t;t_{0},\mathbf{x}_{0})$.

A Lagrangian scalar field $\mathcal{P}(\mathbf{x}_{0})$ is called
objective \cite{truesdell92} if it remains invariant under all Euclidean
observer changes of the form 
\begin{equation}
\mathbf{x}=\mathbf{Q}(t)\mathbf{y}+\mathbf{b}(t),\label{eq:observer change-1}
\end{equation}
with a proper orthogonal tensor $\mathbf{Q}(t)\in\mathrm{SO}(n)$
and a vector $\mathbf{b}(t)$. Specifically, under all such transformations,
the transformed scalar field $\tilde{\mathcal{P}}$ in the $\mathbf{y}$-frame
satisfies $\tilde{\mathcal{P}}(\mathbf{y}_{0})\equiv\mathcal{P}(\mathbf{x}_{0}).$

Consider a material curve, \textbf{$\boldsymbol{\gamma}(t;s)=\mathbf{F}_{t_{0}}^{t}\left(\boldsymbol{\gamma}(t_{0};s)\right)\subset U$},
at time $t$, which has evolved from an initial curve $\boldsymbol{\gamma}(t_{0};s)$
parametrized by\textbf{ }the scalar parameter $s\in\mathbb{R}$. Any
tangent vector, 
\begin{equation}
\boldsymbol{\xi}(t;s)=\partial_{s}\boldsymbol{\gamma}(t;s),\label{eq:tangent vector of material line}
\end{equation}
of \textbf{$\boldsymbol{\gamma}$} then satisfies the evolution equation
\begin{align*}
\dot{\boldsymbol{\xi}}(t;s) & =\partial_{t}\boldsymbol{\xi}(t;s)=\partial_{t}\partial_{s}\boldsymbol{\gamma}(t;s)=\partial_{s}\partial_{t}\boldsymbol{\gamma}(t;s)=\partial_{s}\partial_{t}\mathbf{F}_{t_{0}}^{t}\left(\boldsymbol{\gamma}(t_{0};s)\right)\\
 & =\partial_{s}\mathbf{v}(\boldsymbol{\gamma}(t;s),t)=\boldsymbol{\nabla}\mathbf{v}(\boldsymbol{\gamma}(t;s),t)\partial_{s}\boldsymbol{\gamma}(t;s).
\end{align*}
Therefore, the tangent vector $\boldsymbol{\xi}(t;s)$ of the material
curve $\boldsymbol{\gamma}(t;s)$ satisfies the classic equation of
variations \cite{arnold92} 
\begin{equation}
\dot{\boldsymbol{\xi}}=\boldsymbol{\nabla}\mathbf{v}\left(\boldsymbol{x}(t),t\right)\boldsymbol{\xi}\label{eq:eq of variations}
\end{equation}
along the trajectory $\boldsymbol{x}(t)$ with $\boldsymbol{x}(t_{0})=\boldsymbol{\gamma}(t_{0};s).$

\subsection{Objective stretching measures}

We now fix the parameter $s$ along the material curve $\boldsymbol{\gamma}(t;s)$
such that $\boldsymbol{\gamma}(t_{0};s)=\mathbf{\boldsymbol{x}}_{0}$.
We then seek to characterize the time-evolution of the tangent vector
$\left|\boldsymbol{\xi}(t)\right|$ at $\boldsymbol{\gamma}(t;s)$
with an exponent $\lambda$ as 
\begin{equation}
\left|\boldsymbol{\xi}(t)\right|\sim e^{\lambda\left(t-t_{0}\right)}\left|\boldsymbol{\xi}_{0}\right|\label{eq:sigmadef heuristic}
\end{equation}
for $\boldsymbol{\xi}_{0}\neq\mathbf{0}$. This leads us to define
the \emph{averaged stretching exponent} 
\begin{equation}
\lambda{}_{t_{0}}^{t_{N}}(\mathbf{\boldsymbol{x}}_{0},\boldsymbol{\xi}_{0}):=\frac{1}{t_{N}-t_{0}}\log\frac{\left|\boldsymbol{\xi}(t_{N})\right|}{\left|\boldsymbol{\xi}_{0}\right|},\label{eq:lambdadef}
\end{equation}
sometimes called the finite-time Lyapunov exponent associated with
the initial vector $\boldsymbol{\xi}_{0}$ at the initial location
$\mathbf{\boldsymbol{x}}_{0}$ (see, e.g., \cite{ott08}). Under any
Euclidean observer change (\ref{eq:observer change-1}) we obtain
\[
\tilde{\mathbf{\boldsymbol{\xi}}}=\mathbf{Q}^{T}\boldsymbol{\xi},
\]
whose substitution into (\ref{eq:lambdadef}) shows that the Lagrangian
scalar field $\lambda_{t_{0}}^{t_{N}}(\boldsymbol{x}\mathbf{}_{0},\boldsymbol{\xi}_{0})$
is objective, i.e., remains invariant under arbitrary observer changes
of the form (\ref{eq:observer change-1}). We recall that the maximum
of $\lambda>0$ with respect to the direction $\boldsymbol{\xi}_{0}$
is the finite-time Lyapunov exponent (FTLE) associated with the initial
point $\boldsymbol{x}\mathbf{}_{0}$ by definition \cite{haller15}.

If the instantaneous growth exponent satisfies $\lambda<0$ at time
$t$ in a two-dimensional incompressible flow, then a normal vector
$\boldsymbol{\xi}^{\perp}(t;s)$ to $\boldsymbol{\xi}(t;s)$ at the
same point of the trajectory will have a growth exponent equal to
$\left|\lambda\right|>0$. Similarly, in three-dimensional incompressible
flows, $\lambda<0$ implies that the instantaneous growth exponent
of the \emph{area} in a plane normal to the material line $\boldsymbol{\gamma}(t;s)$
is equal to $\left|\lambda\right|>0$. Therefore, $\left|\lambda\right|$
is a measure of the strength of hyperbolicity experienced by a trajectory
in an incompressible flow. To measure the temporal average of $\left|\lambda\right|$
along the trajectory $\boldsymbol{x}(t)$, we define the \emph{averaged
hyperbolicity strength} 
\begin{equation}
\bar{\lambda}{}_{t_{0}}^{t_{N}}(\mathbf{\boldsymbol{x}}_{0},\boldsymbol{\xi}_{0}):=\frac{1}{t_{N}-t_{0}}\int_{t_{0}}^{t_{N}}\left|\frac{d}{dt}\log\frac{\left|\boldsymbol{\xi}(t)\right|}{\left|\boldsymbol{\xi}_{0}\right|}\right|\,dt,\label{eq:lambdabardef}
\end{equation}
where $\boldsymbol{x}(t_{0})=\mathbf{\boldsymbol{x}}_{0}$. While
the quantity $\lambda{}_{t_{0}}^{t_{N}}$ measures the average net
stretching (including shrinkage) in the direction of $\boldsymbol{\xi}(t)$
over the time interval $[t_{0},t_{N}]$ , the quantity $\bar{\lambda}{}_{t_{0}}^{t_{N}}$
accounts for all (positive) stretching experienced either parallel
or normal to $\boldsymbol{\xi}(t)$. For this reason, $\bar{\lambda}{}_{t_{0}}^{t_{N}}$
is a measure of the average strength of hyperbolicity experienced
by a trajectory of an incompressible flow. By the same considerations
that we have applied to $\lambda{}_{t_{0}}^{t_{N}}(\mathbf{\boldsymbol{x}}_{0},\boldsymbol{\xi}_{0})$,
the scalar field $\bar{\lambda}{}_{t_{0}}^{t_{N}}(\mathbf{\boldsymbol{x}}_{0},\boldsymbol{\xi}_{0})$
is also objective.

\subsection{Objective rotation measures}

To introduce our second class of trajectory-based diagnostic tool,
we normalize the material tangent vector $\boldsymbol{\xi}(t)$ defined
in (\ref{eq:tangent vector of material line}) to the unit tangent
vector 
\[
\mathbf{e}(t)=\frac{\boldsymbol{\xi}(t)}{\left|\boldsymbol{\xi}(t)\right|}
\]
of the material line $\boldsymbol{\gamma}(t;s)$; we will denote by
$\mathbf{e}_{0}:=\mathbf{e}(t_{0})=\boldsymbol{\xi}_{0}/\left|\boldsymbol{\xi}_{0}\right|$
the initial position of $\mathbf{e}(t)$ at time $t_{0}$.

Under any Euclidean observer change (\ref{eq:observer change-1})
we then obtain 
\begin{equation}
\tilde{\mathbf{e}}(t)=\mathbf{Q}^{T}(t)\mathbf{e}(t).\label{eq:transformation of unit material vector}
\end{equation}
Over the time interval $[t_{0},t_{N}],$ the net rotation angle $\alpha_{t_{0}}^{t_{N}}(\mathbf{x}_{0},\mathbf{e}_{0})$
per unit time experienced by $\boldsymbol{\xi}_{0}$ along the trajectory
starting from\textbf{ }$\mathbf{x}_{0}=\boldsymbol{\gamma}(t_{0};s)$
can be computed as 
\begin{equation}
\alpha_{t_{0}}^{t_{N}}(\mathbf{x}_{0},\mathbf{e}_{0})=\frac{1}{t_{N}-t_{0}}\cos^{-1}\left\langle \mathbf{e}(t_{0}),\mathbf{e}(t_{N})\right\rangle .\label{eq:alphadef-1}
\end{equation}

Formula (\ref{eq:transformation of unit material vector}) shows that
$\alpha_{t_{0}}^{t}$ is not an objective scalar because $\mathbf{e}_{0}=\mathbf{e}(t_{0})$
is rotated by $\mathbf{Q}^{T}(t_{0})$ under an observer change (\ref{eq:observer change-1}),
whereas $\mathbf{e}(t_{N})$ is rotated by $\mathbf{Q}^{T}(t_{N}).$
Still, rotations commute in two dimensions, so if we have $\alpha_{t_{0}}^{t}(\mathbf{x}_{01},\mathbf{e}_{01})=\alpha_{t_{0}}^{t}(\mathbf{x}_{02},\mathbf{e}_{02})$
for two different initial positions $\mathbf{x}_{01}$ and $\mathbf{x}_{02}$
with corresponding initial material tangent unit vectors $\mathbf{e}_{01}$
and $\mathbf{e}_{02}$, then we also have $\alpha_{t_{0}}^{t}(\mathbf{y}_{01},\mathbf{\tilde{e}}_{01})=\alpha_{t_{0}}^{t}(\mathbf{y}_{02},\mathbf{\tilde{e}}_{02})$
for their transformed counterparts. This is because the angle between
$\mathbf{e}_{1}(t_{0})$ and $\mathbf{e}_{1}(t_{N})$ will be altered
by the spatially independent rotation tensors $\mathbf{Q}^{T}(t_{0})$
and $\mathbf{Q}^{T}(t_{N})$ in the new frame by the same amount as
the angle between $\mathbf{e}_{2}(t_{0})$ and $\mathbf{e}_{2}(t_{N})$
will be altered by the same rotation tensors. We conclude that the
set of level curves of (\ref{eq:alphadef-1}) is objective (i.e.,
invariant under observer changes), even though the value of $\alpha_{t_{0}}^{t}(\mathbf{x}_{0},\mathbf{e}_{0})$
on them changes from one frame to the other. We stress that the same
statement does not hold in three dimensions, unless the vector field
$\mathbf{v}$ itself is objective. These properties mimic the objectivity
properties of the PRA (\cite{farazmand16}).

We now seek to construct a quantity that objectively characterizes
the rotation of unit vectors in both two and three dimensions. To
this end, we consider the deviation of $\dot{\mathbf{e}}$ from the
domain-average of the local mean velocities of all unit material tangent
vectors emanating from the same point. We first recall the result
from \cite{haller16a} that the local mean angular velocity of material
fibers at a point $\mathbf{x}$ at time $t$ equals 
\begin{equation}
\boldsymbol{\nu}(\mathbf{x},t)=\frac{1}{2}\boldsymbol{\omega}(\mathbf{x},t),\label{eq:nudef}
\end{equation}
with $\boldsymbol{\omega}=\boldsymbol{\nabla}\times\mathbf{v}$ denoting
the vorticity field. We also recall that under a frame change (\ref{eq:observer change-1}),
vorticity transforms into 
\begin{equation}
\tilde{\boldsymbol{\omega}}=\mathbf{Q}^{T}\left(\boldsymbol{\omega}-\dot{\mathbf{q}}\right),\label{eq:vorticity transformation formula}
\end{equation}
where $\dot{\mathbf{q}}$ is the vorticity of the frame change defined
by the identity 
\[
\frac{1}{2}\dot{\mathbf{q}}\times\mathbf{e}=\dot{\mathbf{Q}}\mathbf{Q}^{T}\mathbf{e}
\]
for all vectors $\mathbf{e}\in\mathbb{R}^{3}.$

By eq. (\ref{eq:nudef}), the deviation of $\dot{\mathbf{e}}$ from
the spatially averaged local velocities of material unit vectors,
or \emph{relative tangent velocity}, is 
\begin{equation}
\dot{\boldsymbol{\epsilon}}=\dot{\mathbf{e}}-\frac{1}{2}\bar{\boldsymbol{\omega}}(t)\times\mathbf{e}=\dot{\mathbf{e}}-\bar{\mathbf{W}}(t)\mathbf{e},\label{eq:epsilon dot}
\end{equation}
with $\bar{\boldsymbol{\omega}}(t)$ denoting the spatial average
of the vorticity over the domain $U$, and with $\bar{\mathbf{W}}(t)$
denoting the spatial average of the spin tensor 
\begin{equation}
\mathbf{W}=\frac{1}{2}\left[\boldsymbol{\nabla}\mathbf{v}-\left[\boldsymbol{\nabla}\mathbf{v}\right]^{T}\right]=-\mathbf{W}^{T}\label{eq:spin tensor}
\end{equation}
over the same domain. Under the frame change (\ref{eq:observer change-1}),
formula (\ref{eq:vorticity transformation formula}) shows that $\dot{\boldsymbol{\epsilon}}$
transforms as 
\begin{align*}
\dot{\tilde{\boldsymbol{\epsilon}}} & =\dot{\tilde{\mathbf{e}}}-\frac{1}{2}\bar{\tilde{\boldsymbol{\omega}}}\times\tilde{\mathbf{e}}=\mathbf{Q}^{T}\left[\dot{\mathbf{e}}-\dot{\mathbf{Q}}\mathbf{Q}^{T}\mathbf{e}\right]-\frac{1}{2}\mathbf{Q}^{T}\left(\bar{\boldsymbol{\omega}}-\dot{\mathbf{q}}\right)\times\mathbf{Q}^{T}\mathbf{e}\\
 & =\mathbf{Q}^{T}\left[\dot{\mathbf{e}}-\frac{1}{2}\dot{\mathbf{q}}\times\mathbf{e}\right]-\frac{1}{2}\mathbf{Q}^{T}\left(\bar{\boldsymbol{\omega}}-\dot{\mathbf{q}}\right)\times\mathbf{Q}^{T}\mathbf{e}\\
 & =\mathbf{Q}^{T}\left[\dot{\mathbf{e}}-\frac{1}{2}\dot{\mathbf{q}}\times\mathbf{e}-\frac{1}{2}\left(\bar{\boldsymbol{\omega}}-\dot{\mathbf{q}}\right)\times\mathbf{e}\right]\\
 & =\mathbf{Q}^{T}\left[\dot{\mathbf{e}}-\frac{1}{2}\bar{\boldsymbol{\omega}}\times\mathbf{e}\right]=\mathbf{Q}^{T}\dot{\boldsymbol{\epsilon}},
\end{align*}
implying that the relative tangent velocity $\dot{\boldsymbol{\epsilon}}$
of $\boldsymbol{\xi}$ is an objective vector. Consequently, 
\begin{equation}
\bar{\alpha}_{t_{0}}^{t_{N}}(\mathbf{\boldsymbol{x}}_{0},\boldsymbol{\xi}_{0}):=\frac{1}{t_{N}-t_{0}}\int_{t_{0}}^{t_{N}}\left|\dot{\boldsymbol{\epsilon}}(t)\right|dt=\frac{1}{t_{N}-t_{0}}\int_{t_{0}}^{t_{N}}\left|\dot{\mathbf{e}}(t)-\frac{1}{2}\bar{\boldsymbol{\omega}}(t)\times\mathbf{e}(t)\right|dt,\label{eq:alphabardef}
\end{equation}
with $\mathbf{e}(t)=$ $\boldsymbol{\xi}(t)/\left|\boldsymbol{\xi}(t)\right|$,
is an objective measure for the average rotation speed experienced
by the tangent vector $\boldsymbol{\xi}$ during its evolution along
the trajectory starting from $\mathbf{x}_{0}$. Indeed, this rate
of change is entirely due to the rotation of $\boldsymbol{\xi}$,
unaffected by any change in the length of $\boldsymbol{\xi}$. 

We have used the $\bar{\alpha}_{t_{0}}^{t_{N}}$ notation in \eqref{eq:alphabardef}
to point out that this quantity (the averaged norm of an angular velocity)
has the same relation to $\alpha_{t_{0}}^{t_{N}}$ (an average angular
velocity) as the averaged norm of the stretching rate $\bar{\lambda}{}_{t_{0}}^{t_{N}}$
has to the average stretching rate $\lambda{}_{t_{0}}^{t_{N}}$. Unlike
$\lambda{}_{t_{0}}^{t_{N}}$, however, $\alpha_{t_{0}}^{t_{N}}$ is
not objective and that is why the definition of $\bar{\alpha}_{t_{0}}^{t_{N}}$
contains a subtracted mean rotation rate that makes $\bar{\alpha}_{t_{0}}^{t_{N}}$
objective.

\section{Quasi-objective trajectory stretching and rotation in steady flows\label{sec: Steady flows}}

We will call a Lagrangian scalar field $P(\mathbf{x}_{0})$ computed
from $\left\{ \boldsymbol{x}(t_{i})\right\} _{i=0}^{N}$ \emph{quasi-objective
under a condition }\textbf{(A)} if there exists an objective Lagrangian
scalar field $\mathcal{P}(\mathbf{x}_{0})$ such that 
\begin{equation}
P(\mathbf{x}_{0})\approx\mathcal{P}(\mathbf{x}_{0})\label{eq:L=00003D00003D00003D00003D00003D00003DP}
\end{equation}
in any $\mathbf{x}$-frame in which condition\emph{ }(\textbf{A)}
holds. The accuracy of the approximation indicated by the symbol $\approx$
in (\ref{eq:L=00003D00003D00003D00003D00003D00003DP}) depends on
the extent to which the condition (\textbf{A)} is satisfied.

In other words, a frame-dependent scalar field $P(\mathbf{x}_{0})$
is quasi-objective under a condition $(\mathbf{A})$ if it happens
to approximate the same objective scalar field $\mathcal{P}(\mathbf{x}_{0})$
in any frame in which condition $(\mathbf{A})$ is satisfied. $P(\mathbf{x}_{0})$
will generally not approximate an objective field in all frames, as
condition $(\mathbf{A})$ is generally frame-dependent and hence will
not hold in all frames. In the following sections, we will derive
quasi-objective measures of material stretching and rotation with
computable formulas for their corresponding quasi-objectivity condition.

Our first quasi-objectivity conditions for our upcoming derivations
will be the assumption that the velocity field is steady in the current
frame, i.e., 
\begin{description}
\item [{(A1)~~~~~~~~~~~~~~~~~~~~~~~~~~~~~~~~~$\left|\partial_{t}\mathbf{v}(\mathbf{x},t)\right|=0,$}] ~ 
\end{description}
holds. We will remove this assumption in our later extension of our
results to unsteady flows.

Assumption (\textbf{A1)} enables us to rewrite (\ref{eq:velocity field})
as the autonomous dynamical system

\begin{equation}
\dot{\mathbf{x}}=\mathbf{v}(\mathbf{x}).\label{eq:velocity field-1}
\end{equation}
Trajectories of two-dimensional steady flows ($n=2$) coincide with
streamlines and already provide detailed information about coherent
structures without further analysis. For this reason, the present
discussion is mostly of interest for three-dimensional flows ($n=3$).
Although the flow map and trajectories of the autonomous differential
equation depend on the elapsed time $t-t_{0}$ and hence $t_{0}=0$
could be fixed, we still keep the explicit dependence on the current
time $t$ and initial time $t_{0}$ in our formulas to facilitate
their later extension to the unsteady case in which (\textbf{A1})
will no longer be assumed.

\subsection{Trajectory-based approximations of stretching measures}

Consider now the Lagrangian velocity vector\textbf{ 
\[
\boldsymbol{v}(t):=\mathbf{v}(\boldsymbol{x}(t)),
\]
} which satisfies the homogeneous linear evolution equation 
\begin{equation}
\dot{\mathbf{\mathbf{\boldsymbol{v}}}}=\boldsymbol{\nabla}\mathbf{v}(\boldsymbol{x}(t))\mathbf{\boldsymbol{v}}.\label{eq:evolution of velocity vector}
\end{equation}
Note that this formula relies heavily on our assumption (\textbf{A1}),
otherwise the extra term $\partial_{t}\mathbf{v}(\boldsymbol{x}(t),t)$
would appear on the right-hand side, making (\ref{eq:evolution of velocity vector})
an inhomogeneous linear system of differential equations.

A comparison of eqs. (\ref{eq:eq of variations}) and (\ref{eq:evolution of velocity vector})
shows that $\boldsymbol{v}(t)$ evolves as a material tangent vector
in the given coordinate frame. Therefore, for the choice of the initial
material tangent vector $\boldsymbol{\xi}_{0}=\mathbf{\boldsymbol{v}}_{0}:=\mathbf{v}(\boldsymbol{x}_{0})\neq\mathbf{0}$,
the averaged stretching exponent and averaged hyperbolicity strength
can be written as 
\begin{align}
\lambda{}_{t_{0}}^{t_{N}}(\boldsymbol{x}\mathbf{}_{0},\mathbf{\boldsymbol{v}}_{0})= & \frac{1}{t_{N}-t_{0}}\log\frac{\left|\mathbf{v}\left(\boldsymbol{x}(t_{N})\right)\right|}{\left|\mathbf{\boldsymbol{v}}_{0}\right|},\qquad\bar{\lambda}{}_{t_{0}}^{t_{N}}(\boldsymbol{x}\mathbf{}_{0},\mathbf{\boldsymbol{v}}_{0})=\frac{1}{t_{N}-t_{0}}\int_{t_{0}}^{t_{N}}\left|\frac{d}{dt}\log\frac{\left|\mathbf{v}\left(\boldsymbol{x}(t)\right)\right|}{\left|\mathbf{\boldsymbol{v}}_{0}\right|}\right|\,dt.\label{eq:lambda and lambdabar approx}
\end{align}
The right-hand sides of the formulas in (\ref{eq:lambda and lambdabar approx})
give close approximation for the objective expressions (\ref{eq:lambdadef})
and (\ref{eq:lambdabardef}) with $\boldsymbol{\xi}_{0}=\mathbf{\boldsymbol{v}}_{0}$
in the frame where assumption (\textbf{A1}) holds for the velocity
field. In other words, $\lambda{}_{t_{0}}^{t_{N}}(\boldsymbol{x}\mathbf{}_{0},\mathbf{\boldsymbol{v}}_{0})$
and $\bar{\lambda}{}_{t_{0}}^{t_{N}}(\boldsymbol{x}\mathbf{}_{0},\mathbf{\boldsymbol{v}}_{0})$
are quasi-objective under condition (\textbf{A1}).

If discretized trajectory data $\left\{ \boldsymbol{x}(t_{i})\right\} _{i=0}^{N}$
is available for the trajectory $\boldsymbol{x}(t)$, then formulas
(\ref{eq:lambda and lambdabar approx}) yield the following result,
with the trajectory velocities $\dot{\mathbf{\boldsymbol{x}}}(t_{j})$
computed from finite differencing or spline interpolation. 
\begin{thm}
Under assumption (\textbf{A1})\emph{, }the trajectory stretching exponents\emph{
}(TSEs), defined as 
\begin{equation}
\mathrm{TSE}_{t_{0}}^{t_{N}}(\boldsymbol{x}\mathbf{}_{0})=\frac{1}{t_{N}-t_{0}}\log\frac{\left|\dot{\mathbf{\boldsymbol{x}}}(t_{N})\right|}{\left|\dot{\mathbf{\boldsymbol{x}}}(t_{0})\right|},\qquad\overline{\mathrm{TSE}}_{t_{0}}^{t_{N}}(\boldsymbol{x}\mathbf{}_{0})=\frac{1}{t_{N}-t_{0}}\sum_{i=0}^{N-1}\left|\log\frac{\left|\dot{\mathbf{\boldsymbol{x}}}(t_{i+1})\right|}{\left|\dot{\mathbf{\boldsymbol{x}}}(t_{i})\right|}\right|\label{eq:tajectory-based stretching exponents}
\end{equation}
for all $\dot{\mathbf{\boldsymbol{x}}}(t_{i})\neq\mathbf{0}$, $i=0,\ldots,N$,
are quasi-objective measures of trajectory stretching and hyperbolicity
strength. 
\end{thm}

The following remarks are in order regarding Theorem 1: 
\begin{enumerate}
\item While two-dimensional steady flows can already be fully understood
from streamlines generated by trajectory plots, the analysis of three-dimensional
steady flows benefits from the $\mathrm{TSE}$ diagnostics. 
\item The $\mathrm{TSE}$ metrics are only quasi-objective, not objective.
One manifestation of this is that they depend on trajectory velocities
which are not objective. That said, the $\mathrm{TSE}$ metrics give
the correct stretching of velocity vectors in case those vectors evolve
materially in a given frame, which is guaranteed by condition (\textbf{A1}). 
\item The $\mathrm{TSE}$ metrics can generally be used to highlight hyperbolic
invariant manifolds of an autonomous vector field. That vector field
does not have to be a fluid velocity field and hence, unlike fluid
velocity fields, it may be objective. For such objective autonomous
vector fields, the $\mathrm{TSE}$ metrics are objective, not only
quasi-objective, given that assumption (\textbf{A1}) is satisfied
in any observer frame. Examples of objective autonomous vector fields
include the active barrier vector fields $\mathbf{x}^{\prime}=\mathbf{w}(\mathbf{x};t)$
derived in \cite{haller20}, with the prime denoting differentiation
with respect to a parameter along a barrier curve. With respect to
this parameter, the vector field $\mathbf{w}(\mathbf{x};t)$ is autonomous.
Under observer changes of the form \eqref{eq:observer change-1},
$\mathbf{w}(\mathbf{x};t)$ remains autonomous and transforms objectively
as $\mathbf{y}^{\prime}=\tilde{\mathbf{w}}(\mathbf{y};t)=\mathbf{Q}^{T}(t)\mathbf{w}(\mathbf{x};t)$.
Invariant manifolds of $\mathbf{w}(\mathbf{x};t)$ are shown in \cite{haller20}
to coincide with time $t$ positions of objective material barriers
to vorticity and momentum in the underlying fluid flow. In Section
\ref{subsec:Momentum-barriers-in ABC flow}, we show the application
of the $\mathrm{TSE}$ metrics to the computation of objective momentum
transport barriers in the unsteady ABC flow.
\item By its construction, the $\mathrm{TSE}$ field is expected to reproduce
features of the FTLE field but without the computational burdens associated
with the FTLE field. Indeed, unlike the FTLE, the $\mathrm{TSE}$
can be computed even for a single trajectory. 
\item For $N=1$, ridges (or trenches) of the $\mathrm{TSE}$ yield single-trajectory-based
approximation for hyperbolic (or parabolic) objective Eulerian coherent
structures (or OECS, \cite{serra16}), which are the instantaneous
limits of hyperbolic (or parabolic) LCSs. 
\end{enumerate}

\subsection{Trajectory-based approximations of rotation measures}

Consider now the normalized Lagrangian velocity vector 
\[
\mathbf{e}_{\mathbf{\mathbf{\boldsymbol{v}}}}(t)=\frac{\mathbf{\boldsymbol{v}}(t)}{\left|\mathbf{\boldsymbol{v}}(t)\right|}.
\]
Equation \ref{eq:epsilon dot} shows that if $\frac{1}{2}\bar{\boldsymbol{\omega}}(t)\times\mathbf{e}$
is much smaller in norm than $\dot{\mathbf{e}}$, then $\dot{\boldsymbol{\epsilon}}$
is approximately equal to $\dot{\mathbf{e}}$. Therefore, if assumption
(\textbf{A1}) holds (thus $\mathbf{e}_{\mathbf{\mathbf{\boldsymbol{v}}}}(t)$
evolves as a material unit vector $\mathbf{e}(t))$ and, in addition, 
\begin{description}
\item [{(A2)~~~~~~~~~~~~~~~~~~~~~~~~~~~~~~~~~$\left|\frac{1}{2}\bar{\boldsymbol{\omega}}(t)\times\mathbf{\boldsymbol{e}_{v}}(t)\right|\ll\left|\dot{\mathbf{e}}_{\mathbf{v}}(t)\right|$}] ~ 
\end{description}
holds in the given frame, then the relative tangent velocity 
\[
\dot{\boldsymbol{\epsilon}}_{\mathbf{\mathbf{\boldsymbol{v}}}}=\dot{\mathbf{e}}_{\mathbf{\mathbf{\boldsymbol{v}}}}-\frac{1}{2}\bar{\boldsymbol{\omega}}\times\mathbf{e}_{\mathbf{\mathbf{\boldsymbol{v}}}}
\]
of $\mathbf{\boldsymbol{v}}(t)$ satisfies 
\begin{align*}
\left|\dot{\boldsymbol{\epsilon}}_{\mathbf{\mathbf{\boldsymbol{v}}}}\right| & \approx\left|\dot{\mathbf{e}}_{\mathbf{\mathbf{\boldsymbol{v}}}}\right|.
\end{align*}
Note that assumption (\textbf{A2}) requires the Lagrangian accelerations
to dominate the angular acceleration of the trajectory induced by
the spatial mean vorticity. Consequently, if assumptions (\textbf{A1})
and (\textbf{A2}) hold in the frame in which the trajectory data $\boldsymbol{x}(t)$
is given, then 
\begin{equation}
\bar{\alpha}_{t_{0}}^{t_{N}}(\mathbf{\boldsymbol{x}}_{0},\boldsymbol{v}_{0}):=\frac{1}{t_{N}-t_{0}}\int_{t_{0}}^{t_{N}}\left|\dot{\mathbf{e}}_{\mathbf{\mathbf{\boldsymbol{v}}}}(t)\right|dt\label{eq:alphabardef-1}
\end{equation}
approximates the average of $\left|\dot{\boldsymbol{\epsilon}}_{\mathbf{\mathbf{\boldsymbol{v}}}}\right|$.

To see the physical meaning of $\bar{\alpha}_{t_{0}}^{t_{N}}(\mathbf{\boldsymbol{x}}_{0},\boldsymbol{v}_{0})$
more specifically, we set $t_{0}=t$ and $t_{N}=t+\delta$ in formula
(\ref{eq:alphadef-1}) to obtain 
\begin{equation}
\cos\left(\alpha_{t}^{t+\delta}\delta\right)=\left\langle \mathbf{e}_{\mathbf{\mathbf{\boldsymbol{v}}}}(t),\mathbf{e}_{\mathbf{\mathbf{\boldsymbol{v}}}}(t+\delta)\right\rangle \label{eq:angle formula delta}
\end{equation}
for the angle $\alpha_{t}^{t+\delta}\delta$ between $\mathbf{e}_{\mathbf{\mathbf{\boldsymbol{v}}}}(t)$
and $\mathbf{e}_{\mathbf{\mathbf{\boldsymbol{v}}}}(t+\delta)$. Differentiating
(\ref{eq:angle formula delta}) with respect to $\delta$ leads to
the expression 
\[
-\left|\alpha_{t}^{t+\delta}\right|^{2}\delta+\mathcal{O}(\delta^{2})=\left\langle \mathbf{e}_{\mathbf{\mathbf{\boldsymbol{v}}}}(t),\dot{\mathbf{e}}_{\mathbf{\mathbf{\boldsymbol{v}}}}(t+\delta)\right\rangle ,
\]
where we have assumed that $\lim_{\delta\to0}\left|\alpha_{t}^{t+\delta}\right|$
exists and is bounded. Dividing both sides by $\left(-\delta\right)$
and taking the $\delta\to0$ limit gives 
\begin{equation}
\left|\alpha_{t}^{t}\right|^{2}=\lim_{\delta\to0}\left[-\left\langle \mathbf{e}_{\mathbf{\mathbf{\boldsymbol{v}}}}(t),\ddot{\mathbf{e}}_{\mathbf{\mathbf{\boldsymbol{v}}}}(t+\delta)\right\rangle \right]=\left|\dot{\mathbf{e}}_{\mathbf{\mathbf{\boldsymbol{v}}}}(t)\right|^{2},\label{eq:formula  for alpha dot}
\end{equation}
where the last equality follows from differentiating the identity
$\left\langle \mathbf{e}(t),\mathbf{e}(t)\right\rangle \equiv1$ twice
with respect to $t$. Formula (\ref{eq:formula  for alpha dot}) shows
that $\lim_{\delta\to0}\left|\alpha_{t}^{t+\delta}\right|$ indeed
exists and is bounded, as assumed. Using this formula in \eqref{eq:alphabardef-1}
then gives
\begin{equation}
\bar{\alpha}_{t_{0}}^{t_{N}}(\mathbf{\boldsymbol{x}}_{0},\boldsymbol{v}_{0})=\frac{1}{t_{N}-t_{0}}\int_{t_{0}}^{t_{N}}\left|\alpha_{t}^{t}\right|dt,\label{eq:theory for TAV}
\end{equation}
showing that $\bar{\alpha}_{t_{0}}^{t_{N}}(\mathbf{\boldsymbol{x}}_{0},\boldsymbol{v}_{0})$
is the time-average of $\left|\alpha_{t}^{t}\right|$. If only discretized
trajectory data $\left\{ \boldsymbol{x}(t_{i})\right\} _{i=1}^{N}$
is available, then a discretized approximation of (\ref{eq:theory for TAV})
leads to the following result. 
\begin{thm}
Under assumptions (\textbf{A1}) and (\textbf{A2}), the trajectory
angular velocity, defined as 
\begin{equation}
\overline{\mathrm{TRA}}_{t_{0}}^{t_{N}}(\boldsymbol{x}\mathbf{}_{0})=\frac{1}{t_{N}-t_{0}}\sum_{i=0}^{N-1}\cos^{-1}\frac{\left\langle \dot{\mathbf{\boldsymbol{x}}}(t_{i}),\dot{\mathbf{\boldsymbol{x}}}(t_{i+1})\right\rangle }{\left|\dot{\mathbf{\boldsymbol{x}}}(t_{i})\right|\left|\dot{\mathbf{\boldsymbol{x}}}(t_{i+1})\right|}\label{eq:tajectory-based stretching exponent-1}
\end{equation}
for all $\dot{\mathbf{\boldsymbol{x}}}(t_{i})\neq\mathbf{0}$, $i=0,\ldots,N$,
is a quasi-objective measure of total trajectory rotation. 
\end{thm}

Indeed, the $i^{th}$ term of the sum in (\ref{eq:tajectory-based stretching exponent-1})
measures the angle between the velocity vectors $\dot{\mathbf{\boldsymbol{x}}}(t_{i})$
and $\dot{\mathbf{\boldsymbol{x}}}(t_{i+1})$, returning always a
positive value to account for the modulus sign in (\ref{eq:theory for TAV}).
As already noted, the trajectory velocities $\dot{\mathbf{\boldsymbol{x}}}(t_{i})$
can be computed from finite differencing or spline interpolation performed
on the subsequent trajectory positions $\mathbf{\boldsymbol{x}}(t_{i})$.

We make the following remarks regarding Theorem 2: 
\begin{enumerate}
\item For the objective autonomous vector fields mentioned in Remark 3 after
Theorem 1, the unit vector field $\mathbf{e}_{\boldsymbol{v}}(t)$
is already objective and hence one does not need to use its modified
version, $\boldsymbol{\epsilon}_{\boldsymbol{v}}(t)$. As a consequence,
assumption (\textbf{A2}) is no longer necessary for such vector fields
and $\mathrm{\mathrm{\overline{\mathrm{TRA}}}}$ is objective without
further assumptions (see \cite{haller20} for examples of such vector
fields). For any $\mathbf{v}(\mathbf{x})$, the net rotation speed
$\alpha_{t_{0}}^{t_{N}}$ defined in (\ref{eq:alphadef-1}) is also
an objective vector field and can be approximated from trajectories
as 
\begin{equation}
\mathrm{TRA}_{t_{0}}^{t_{N}}(\boldsymbol{x}\mathbf{}_{0})=\frac{1}{t_{N}-t_{0}}\cos^{-1}\frac{\left\langle \dot{\mathbf{\boldsymbol{x}}}(t_{0}),\dot{\mathbf{\boldsymbol{x}}}(t_{N})\right\rangle }{\left|\dot{\mathbf{\boldsymbol{x}}}(t_{0})\right|\left|\dot{\mathbf{\boldsymbol{x}}}(t_{N})\right|}\label{eq:objective alpha}
\end{equation}
in both two and three dimensions (see Section \ref{subsec:Momentum-barriers-in ABC flow}
for an application). 
\item Assumption (\textbf{A2}) will automatically hold for any trajectory
with nonzero acceleration if $\bar{\boldsymbol{\omega}}(t)\approx\mathbf{0}$.
The mean vorticity $\bar{\boldsymbol{\omega}}(t)$ has been found
to vanish, up to a very good approximation, in large enough ocean
regions \cite{haller16a,abernathey18,beronvera19}. 
\item By their construction, the $\overline{\mathrm{TRA}}$ and $\mathrm{TRA}$
fields are expected to reproduce features of the PRA field but without
the computational challenges of the PRA field. Indeed, unlike the
PRA, the $\overline{\mathrm{TRA}}$ and $\mathrm{TRA}$ can be computed
even for a single trajectory. 
\item For $N=1$, closed level curves of the $\text{\ensuremath{\mathrm{\overline{\mathrm{TRA}}}}}$
yield single-trajectory-based approximation for elliptic OECS \cite{serra16},
the instantaneous limits of elliptic LCSs. 
\end{enumerate}
In order to verify assumption (\textbf{A2}), one can compute $\ddot{\boldsymbol{x}}(t)$
by finite-differencing as we already noted in the previous section.
In addition, one needs an estimate for the averaged vorticity $\bar{\boldsymbol{\omega}}(t)$,
which may either be generally known for the underlying velocity field
or needs to be estimated from the available trajectory data (see,
e.g., \cite{righi01}).

\section{Quasi-objective coherence measures for unsteady flows\label{sec: Unsteady flows}}

We now extend the results of the previous section to unsteady flows.
The idea is to consider trajectory motion in the extended phase space
of positions and times, in which trajectories are always governed
by a steady velocity field. Each trajectory is a material curve in
the extended phase space. One may, therefore, track the relative stretching
and rotation of the tangent vectors of such curves using the measures
we introduced in the previous section for steady flows. Apart from
the work of \cite{banko19} on an extension of the $Q$-criterion,
coherent structures have apparently not been studied in the extended
phase space.

The quantities $\mathbf{x}$, $t$ and $\mathbf{v}$ generally have
physical units but our upcoming discussion requires non-dimensionalized
variables. To this end, we select a dimensional characteristic length
$L>0$ and a dimensional characteristic time $T>0$ for the flow defined
by (\ref{eq:velocity field}), and introduce the new quantities 
\begin{equation}
\mathbf{y}:=\frac{\mathbf{x}}{L},\quad\tau:=\tau_{0}+\frac{t-t_{0}}{T},\quad\mathrm{v}_{0}:=\frac{L}{T},\quad\mathbf{u}(\mathbf{y},\tau):=\mathbf{u}\left(L\mathbf{y},t_{0}+T\left(\tau-\tau_{0}\right)\right),\label{eq:non-dimensionalization}
\end{equation}
where $\mathbf{y}$ is a non-dimensional position variable, $\tau$
is a non-dimensional time, $\mathrm{v}_{0}$ is a characteristic (dimensional)
velocity for the velocity field\textbf{ $\mathbf{v}(\mathbf{x},t)$
}and $\mathbf{u}(\mathbf{y},\tau)$ is the non-dimensionalized velocity
field. If the variables in (\ref{eq:velocity field}) were non-dimensional
to begin with, then one can simply set $L=T=\mathrm{v}_{0}=1$, $\tau=t$,
\textbf{$\mathbf{y=}\mathbf{x}$ }and $\mathbf{u}\equiv\mathbf{v}$
in all our formulas below.

The non-dimensionalized trajectories satisfy the differential equation
\begin{equation}
\mathbf{y}^{\prime}=\mathbf{u}(\mathbf{y},\tau).\label{eq:nondimensionalized velocity field}
\end{equation}
with $\mathbf{y}\in V\subset\mathbb{R}^{n}$ and $\tau\in[\tau_{0},\tau_{N}].$
We introduce the extended phase space 
\[
\mathcal{V}=\left\{ \mathbf{Y=}\left(\begin{array}{c}
\mathbf{y}\\
z
\end{array}\right):\mathbf{y}\in V,\,\,z\in\mathbb{R}\right\} ,
\]
on which (\ref{eq:velocity field}) becomes an autonomous dynamical
system of the form 
\begin{equation}
\mathbf{Y}^{\prime}=\mathbf{U}(\mathbf{Y}),\qquad\mathbf{U}(\mathbf{Y}):=\left(\begin{array}{c}
\mathbf{u}(\mathbf{y},z)\\
1
\end{array}\right),\qquad\mathbf{Y\in}\mathcal{V}.\label{eq:extended dynamical system}
\end{equation}
Note that without non-dimensionalization, different components of
the extended quantities $\mathbf{Y}$ and $\mathbf{U}$ would have
different physical units.

The autonomous dynamical system (\ref{eq:extended dynamical system})
is of the general form (\ref{eq:velocity field}) with $n=3$ (planar
flows) or $n=4$ (three-dimensional flows). An extended material curve
in this system is of the form 
\begin{equation}
\boldsymbol{\Gamma}(\tau;s)=\left(\begin{array}{c}
\boldsymbol{\gamma}(\tau;s)\\
\gamma_{z}(s)
\end{array}\right)=\left(\begin{array}{c}
\mathbf{F}_{\tau_{0}}^{\tau}\left(\boldsymbol{\gamma}(\tau_{0};s)\right)\\
s
\end{array}\right).\label{eq:extended material curve}
\end{equation}
The initial tangent vector $\boldsymbol{\Xi}_{0}=\partial_{s}\boldsymbol{\Gamma}(\tau_{0};s)$
to this material curve evolves under the extended equation of variations
\begin{equation}
\boldsymbol{\Xi}^{\prime}=\boldsymbol{\partial}_{Y}\mathbf{U}(\mathbf{Y}(\tau))\boldsymbol{\Xi}.\label{eq:extended equation of variation}
\end{equation}
The relationship between the extended material tangent vector $\boldsymbol{\Xi}(\tau)$
and the tangent vector $\boldsymbol{\xi}(t)$ of a material curve
$\boldsymbol{\gamma}(t;s)$ advected by the original velocity field
$\mathbf{v}(\mathbf{x},t)$ is 
\begin{equation}
\boldsymbol{\Xi}(\tau)=\left(\begin{array}{c}
\frac{T}{L}\boldsymbol{\xi}(t)\\
1
\end{array}\right)=\frac{1}{\mathrm{v}_{0}}\left(\begin{array}{c}
\boldsymbol{\xi}(t)\\
\mathrm{v}_{0}
\end{array}\right),\label{eq:extended material tangent vector}
\end{equation}
with $\mathrm{v}_{0}$ denoting the characteristic velocity introduced
in the non-dimensionalization (\ref{eq:non-dimensionalization}).

We can apply the general Lagrangian stretching and rotation measures
introduced in Section \ref{sec:Objective-Lagrangian-measures} to
track the stretching and rotation of $\boldsymbol{\Xi}(\tau)$. As
we have seen, these measures, $\lambda{}_{\tau_{0}}^{\tau_{N}}(\boldsymbol{Y}\mathbf{}_{0},\boldsymbol{\Xi}_{0})$,
$\bar{\lambda}{}_{\tau_{0}}^{\tau_{N}}(\boldsymbol{Y}\mathbf{}_{0},\boldsymbol{\Xi}_{0})$
and $\bar{\alpha}{}_{\tau_{0}}^{\tau_{N}}(\boldsymbol{Y}\mathbf{}_{0},\boldsymbol{\Xi}_{0})$,
will remain invariant under all time-dependent, formally extended
Euclidean transformations 
\begin{equation}
\mathbf{Y}=\boldsymbol{\mathcal{Q}}(\tau)\tilde{\mathbf{Y}}+\mathbf{B}(\tau),\quad\boldsymbol{\mathcal{Q}}(\tau)=\boldsymbol{\mathcal{Q}}^{-T}(\tau)\in SO(n+1),\label{eq:observer change in extended phase space}
\end{equation}
of the extended phase space. In particular, the extended stretching
and rotation measures for $\boldsymbol{\Xi}(\tau)$ remain invariant
under the physically relevant subset of the transformations (\ref{eq:observer change in extended phase space}),
\[
\boldsymbol{\mathcal{Q}}(\tau)=\left(\begin{array}{cc}
\mathbf{Q}(\tau) & \mathbf{0}\\
\mathbf{0}^{T} & 1
\end{array}\right),\quad\mathbf{B}(\tau)=\left(\begin{array}{c}
\mathbf{b}(\tau)\\
0
\end{array}\right),
\]
which represent physical observer changes for the non-dimensionalized
system (\ref{eq:nondimensionalized velocity field}).

Assumption (\textbf{A1) }is always satisfied for the autonomous extended
velocity field (\ref{eq:extended dynamical system}). To evaluate
assumption (\textbf{A2}), we denote the normalized extended velocity
vector and extended averaged spin tensor, respectively, by 
\[
\mathbf{E}_{\mathbf{U}}:=\frac{\mathbf{U}}{\left|\mathbf{U}\right|},\qquad\overline{\boldsymbol{\mathcal{W}}}:=\left(\begin{array}{cc}
\overline{\mathbf{W}}_{\mathbf{u}} & \mathbf{0}\\
\mathbf{0}^{T} & 0
\end{array}\right)=-\overline{\boldsymbol{\mathcal{W}}}^{T},\qquad\mathbf{W}_{\mathbf{u}}=\frac{1}{2}\left[\boldsymbol{\partial}_{\mathbf{y}}\mathbf{u}-\left(\boldsymbol{\boldsymbol{\partial}_{\mathbf{y}}}\mathbf{u}\right)^{T}\right]
\]
with bar denoting spatial averaging over the domain $V$. In terms
of these quantities, assumption (\textbf{A2}) applied to the extended
system takes the form $\left|\overline{\boldsymbol{\mathcal{W}}}\mathbf{E}_{\mathbf{U}}\right|\ll\left|\mathbf{\mathbf{E}_{\mathbf{U}}}^{\prime}\right|$,
or, equivalently, $\left|\bar{\mathbf{W}}\mathbf{e}_{\mathbf{v}}\right|\ll\left|\dot{\mathbf{e}}_{\mathbf{v}}\right|$,
which is just the original form of assumption (\textbf{A2}) in dimensional
coordinates. Theorems 1 and 2 then become applicable in the present
setting, without the requirement (\textbf{A1}), but have to be stated
for the extended Lagrangian velocity 
\[
\mathbf{Y}^{\prime}(\tau_{i})=\left(\begin{array}{c}
\frac{T}{L}\mathbf{v}(\mathbf{x}(t_{i}),t_{i})\\
1
\end{array}\right)=\left(\begin{array}{c}
\frac{1}{\mathrm{v}_{0}}\mathbf{v}(\mathbf{x}(t_{i}),t_{i})\\
1
\end{array}\right)
\]
Substituting these fields into formulas (\ref{eq:tajectory-based stretching exponents})
and (\ref{eq:tajectory-based stretching exponent-1}), we can summarize
the main results for unsteady flows as follows. 
\begin{thm}
(i) The extended trajectory stretching exponents\emph{ }(TSEs), defined
as 
\begin{equation}
\mathrm{TSE}_{t_{0}}^{t_{N}}(\boldsymbol{x}\mathbf{}_{0};\mathrm{v}_{0})=\frac{1}{t_{N}-t_{0}}\log\sqrt{\frac{\left|\dot{\mathbf{x}}(t_{N})\right|^{2}+\mathrm{v}_{0}^{2}}{\left|\dot{\mathbf{x}}(t_{0})\right|^{2}+\mathrm{v}_{0}^{2}}},\qquad\overline{\mathrm{TSE}}_{t_{0}}^{t_{N}}(\boldsymbol{x}\mathbf{}_{0};\mathrm{v}_{0})=\frac{1}{t_{N}-t_{0}}\sum_{i=0}^{N-1}\left|\log\sqrt{\frac{\left|\dot{\mathbf{x}}(t_{i+1})\right|^{2}+\mathrm{v}_{0}^{2}}{\left|\dot{\mathbf{x}}(t_{i})\right|^{2}+\mathrm{v}_{0}^{2}}}\right|\label{eq:tajectory-based stretching exponents-1}
\end{equation}
are objective measures of trajectory stretching and hyperbolicity
strength in the extended phase space.

(ii) Under assumption (\textbf{A2}), the extended trajectory angular
velocity\emph{ }($\mathrm{\overline{\mathrm{TRA}}}$), defined as

\begin{align}
\mathrm{\overline{\mathrm{TRA}}}_{t_{0}}^{t_{N}}(\boldsymbol{x}\mathbf{}_{0};\mathrm{v}_{0}) & =\frac{1}{t_{N}-t_{0}}\sum_{i=0}^{N-1}\cos^{-1}\frac{\left\langle \dot{\mathbf{\boldsymbol{x}}}(t_{i}),\dot{\mathbf{\boldsymbol{x}}}(t_{i+1})\right\rangle +\mathrm{v}_{0}^{2}}{\sqrt{\left|\dot{\mathbf{\boldsymbol{x}}}(t_{i})\right|^{2}+\mathrm{v}_{0}^{2}}\sqrt{\left|\dot{\mathbf{\boldsymbol{x}}}(t_{i+1})\right|^{2}+\mathrm{v}_{0}^{2}}},\label{eq:eq:tajectory-based rotation angle-1}
\end{align}
is a quasi-objective measure of total trajectory rotation in the extended
phase space. 
\end{thm}

The $\mathrm{TSE}$, $\overline{\mathrm{TSE}}$ and $\mathrm{\overline{\mathrm{TRA}}}$
defined in Theorem 3 differ from their counterparts defined in Theorems
1 \& 2. This difference only arises because the same underlying stretching
and rotation metrics are evaluated on the velocity field $\mathbf{v}$
in Theorems 1 \& 2, whereas they are evaluated on the extended velocity
field $(\mathbf{v},1)$ in Theorem 3. Setting the characteristic velocity
$\mathrm{v}_{0}$ to zero in the formulas of Theorem 3, one recovers
the formulas of Theorems 1 \& 2, which, however, are only valid under
assumption \textbf{(A1)}.

We stress that the TSE measures defined in Theorem 3 are objective
without any particular assumption but in the extended phase space.
This means they always approximate an objective measure of material
stretching for the material curve defined by the extended trajectory
$\left(\mathbf{\boldsymbol{x}}(t),t\right)$. The TSE measures, however,
are not objective in the original phase space of the $\mathbf{x}$
variable and they are only quasi-objective in that phase space under
assumption (\textbf{A1}), as we discussed in Theorem 1. In contrast,
the extended $\overline{\mathrm{TRA}}$ measure defined in Theorem
3 is quasi-objective on the extended phase under assumption (\textbf{A2})
and quasi-objective on the original phase space under assumptions
(\textbf{A1})-(\textbf{A2}), as we discussed in Theorem 2.

\section{Examples}

\subsection{Two-dimensional, unsteady ocean surface velocity (AVISO) data set\label{subsec:2D ocean surface data}}

We first evaluate the proposed quasi-objective, single-particle LCS
diagnostics on a two-dimensional, satellite-altimetry-derived ocean-surface
current product (AVISO) that has been the focus of several coherent
structure studies (see, e.g. \cite{beronvera13,haller16b,hadjighasem16,aksamit20}).
The geostrophic component $\mathbf{v}=(v_{1},v_{2})$ of ocean currents
are calculated from the sea-surface height anomaly via the formulas

\begin{equation}
fv_{2}=\frac{1}{\rho}\frac{\partial p}{\partial x},\qquad fv_{1}=-\frac{1}{\rho}\frac{\partial p}{\partial y},\qquad\frac{1}{\rho}\frac{\partial p}{\partial z}=-g,
\end{equation}
where $\rho$ is the density of water, $p$ is the pressure as calculated
from the sea-surface height, $g$ is the constant of gravity, and
$f$ is the Coriolis parameter. A global daily-gridded version of
this data is freely available from the Copernicus Marine Environment
Monitoring Service.

We first verify the minimal differences between the objective continuous
and the quasi-objective discrete formulations of the stretching and
rotation metrics developed in Section \ref{sec: Unsteady flows}.
Our analysis focuses on the Agulhas leakage region 
\begin{equation}
U=\left\{ (x,y)\in\left[-2.5^{\circ},5^{\circ}\right]\times\left[-40^{\circ},-30^{\circ}\right]\right\} \label{eq:U for AVISO}
\end{equation}
where strong hyperbolic and elliptic coherent structures have been
previously identified. Our default rectangular initial grid for generating
trajectories in $U$ will contain $250\times250$ points. Using an
integration time of 25 days, we find the characteristic velocity of
$U$

\begin{equation}
\mathrm{v}_{0}=\frac{1}{NM}\sum_{j=1}^{M}\sum_{i=0}^{N-1}\left|\mathbf{v}(\boldsymbol{x}_{j},t_{i})\right|=0.2\text{ m\ensuremath{\cdot}s}^{-1}
\end{equation}
where $j$ is an index over all initial positions $\boldsymbol{x}_{0}\in U$. 

\begin{figure}[t]
\centering \includegraphics[scale=0.45]{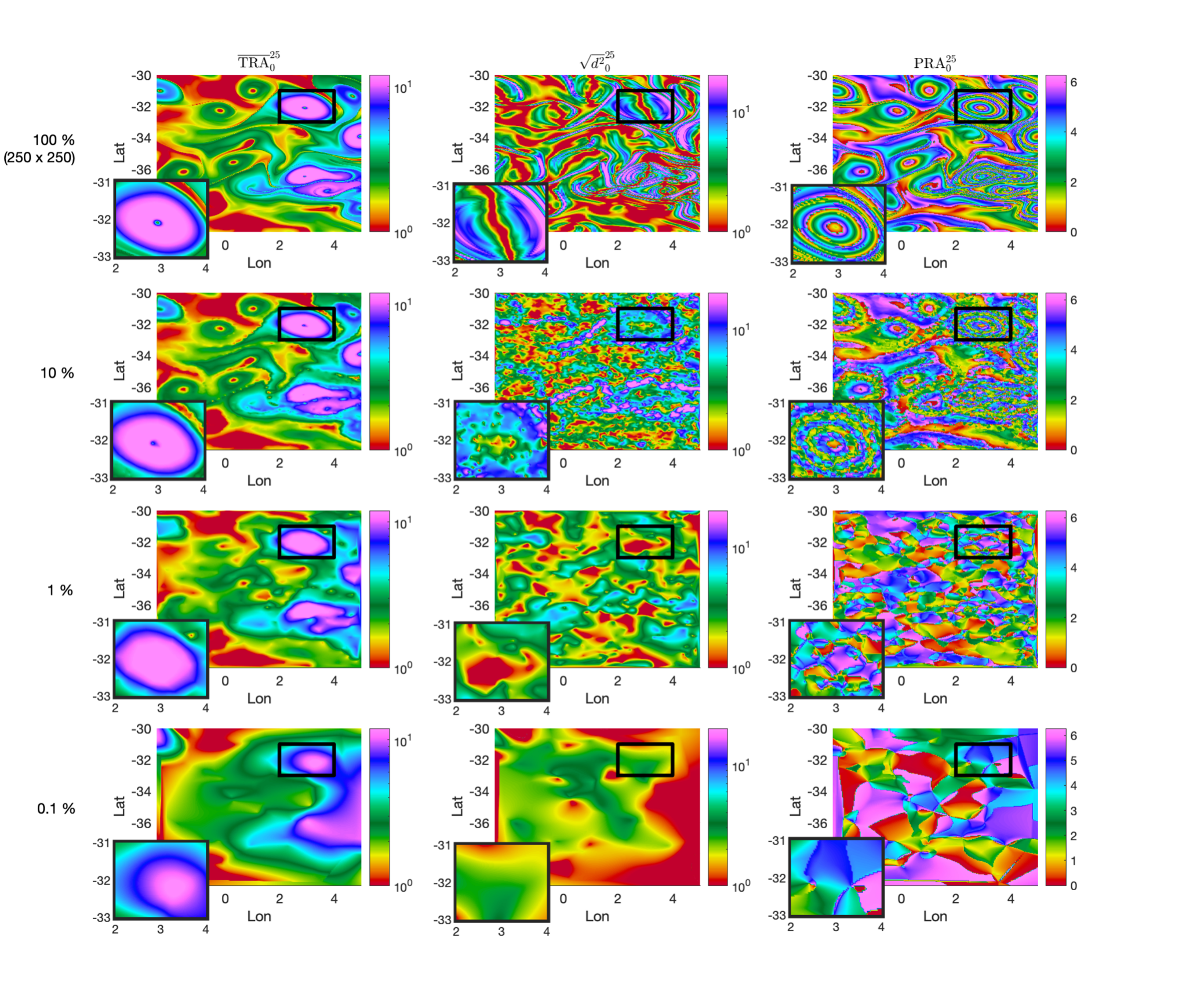} \caption{Rotation metric comparisons for AVISO ocean surface current fields.
The top row shows computations at full resolution with the lower rows
having progressively reduced resolution by randomly subsampling with
fewer trajectories. We display $\sqrt{d^{2}}$ here for clarity.\label{fig:Rotation-metric-comparisons}}
\end{figure}

As already noted in Remark 4 after Theorem 2, condition (\textbf{A2})
is expected to hold for surface currents on large enough ocean domains.
We nevertheless verify (\textbf{A2}) on trajectory data generated
by the AVISO velocity field. Along individual trajectories originating
in $U$, we compute the ratio

\begin{equation}
\delta_{A2}(\boldsymbol{x}\mathbf{}_{0},t_{0},t_{N})=\frac{\int_{t_{0}}^{t_{1}}\left|\frac{1}{2}\bar{\boldsymbol{\omega}}(t)\times\frac{\mathbf{\boldsymbol{v}}(t)}{\left|\mathbf{\boldsymbol{v}}(t)\right|}\right|dt}{\int_{t_{0}}^{t_{1}}\left|\frac{d}{dt}\frac{\mathbf{\boldsymbol{v}}(t)}{\left|\mathbf{\boldsymbol{v}}(t)\right|}\right|dt},\label{eq:A2 ratio}
\end{equation}
which reflects the trajectory-averages of the quantities featured
in assumption (\textbf{A2}). If $\delta_{A2}\ll1$ holds, then (\textbf{A2})
is satisfied on average along the trajectory data serving as a basis
of our analysis. We find that on the domain $U$ defined in (\ref{eq:U for AVISO}),
$\delta_{A2}(\boldsymbol{x}\mathbf{}_{0},0,25)<0.01$ for $98.9\%$
of all initial positions, and $\delta_{A2}(\boldsymbol{x}\mathbf{}_{0},0,25)<0.1$
for all fluid particles. In comparison, on the smaller subdomain $U'=\left[2^{\circ},4^{\circ}\right]\times\left[-33^{\circ},-31^{\circ}\right],$
highlighted as an inset in Fig. \ref{fig:Rotation-metric-comparisons},
the relation $\delta_{A2}(\boldsymbol{x}\mathbf{}_{0},0,25)<0.1$
holds for only $55\%$ of the initial positions and $\delta_{A2}(\boldsymbol{x}\mathbf{}_{0},0,25)<0.01$
holds for only $0.3\%$ of them. This is because this domain contains
only a single coherent structure with substantial average vorticity
that does not represent the near-zero average vorticity of the background
flow surrounding the coherent structures.

Ridges of the forward-time FTLE provide an effective diagnostic for
identifying repelling LCSs \cite{haller15}, but their computation
relies on spatial differentiation of the flow map over a regular grid,
which cannot be carried out from randomly placed single trajectories.
For a fair comparison, we will restrict our calculations of the FTLE
to the same grid resolution as our underlying flow, i.e. we do not
interpolate to a refined grid for spatial differentiation. Another
objective LCS diagnostic we will consider in our comparison is the
single point (squared) relative dispersion \cite{haller00}

\begin{equation}
{d}^{2}(\boldsymbol{x}\mathbf{}_{i0},t_{0},t)=\frac{\left|\boldsymbol{x}\mathbf{}_{i}(t)-\boldsymbol{x}\mathbf{}_{-i}(t)\right|^{2}}{\left|\boldsymbol{x}\mathbf{}_{i0}-\boldsymbol{x}\mathbf{}_{-i0}\right|^{2}},
\end{equation}
where $\boldsymbol{x}{}_{i}$ and $\boldsymbol{x}{}_{-i}$ are trajectory
pairs with initial conditions $\boldsymbol{x}\mathbf{}_{i0}$ and
$\boldsymbol{x}\mathbf{}_{-i0}$ that are initially close at time
$t_{0}$ (e.g., $\left|\boldsymbol{x}\mathbf{}_{i0}-\boldsymbol{x}\mathbf{}_{-i0}\right|=r_{0}\ll1$).
While dividing by the initial distance makes no difference for a uniform
grid of initial conditions, we still include this normalization to
allow for small deviations in $r_{0}$ under a random subsampling
of the initial conditions that mimics real-life sparse trajectory
data. We note that temporal behavior of ensemble averages of relative
dispersion are commonly studied statistics in oceanography \cite{LaCasce08}.

To highlight the significant advantages of $\mathrm{TSE}$ and $\mathrm{TRA}$
(as defined in Theorem 3) on sparse and non-uniform data, we perform
a progressive random subsampling of the grid of initial positions
to degrade the resolution of trajectories in a manner typical of what
is found in experimental data. Computations of $d^{2}$ can accommodate
this subsampling if we choose an $r_{0}$ value for which every initial
position $\boldsymbol{x}\mathbf{}_{i0}$ has a suitable match $\boldsymbol{x}\mathbf{}_{-i0}$.
The FTLE, however, practically requires a regular grid of initial
conditions for the spatial differentiation of the flow map. To this
end, we mimic the standard FTLE computation over an irregular grid
by creating a $C^{1}$-interpolant of final positions, and then computing
the deformation gradient from those interpolations. Alternative methods
for unstructured meshes have been suggested for FTLE but are not the
focus of this study (see e.g. \cite{lekien10}).

\begin{figure}[t]
\centering \includegraphics[scale=0.45]{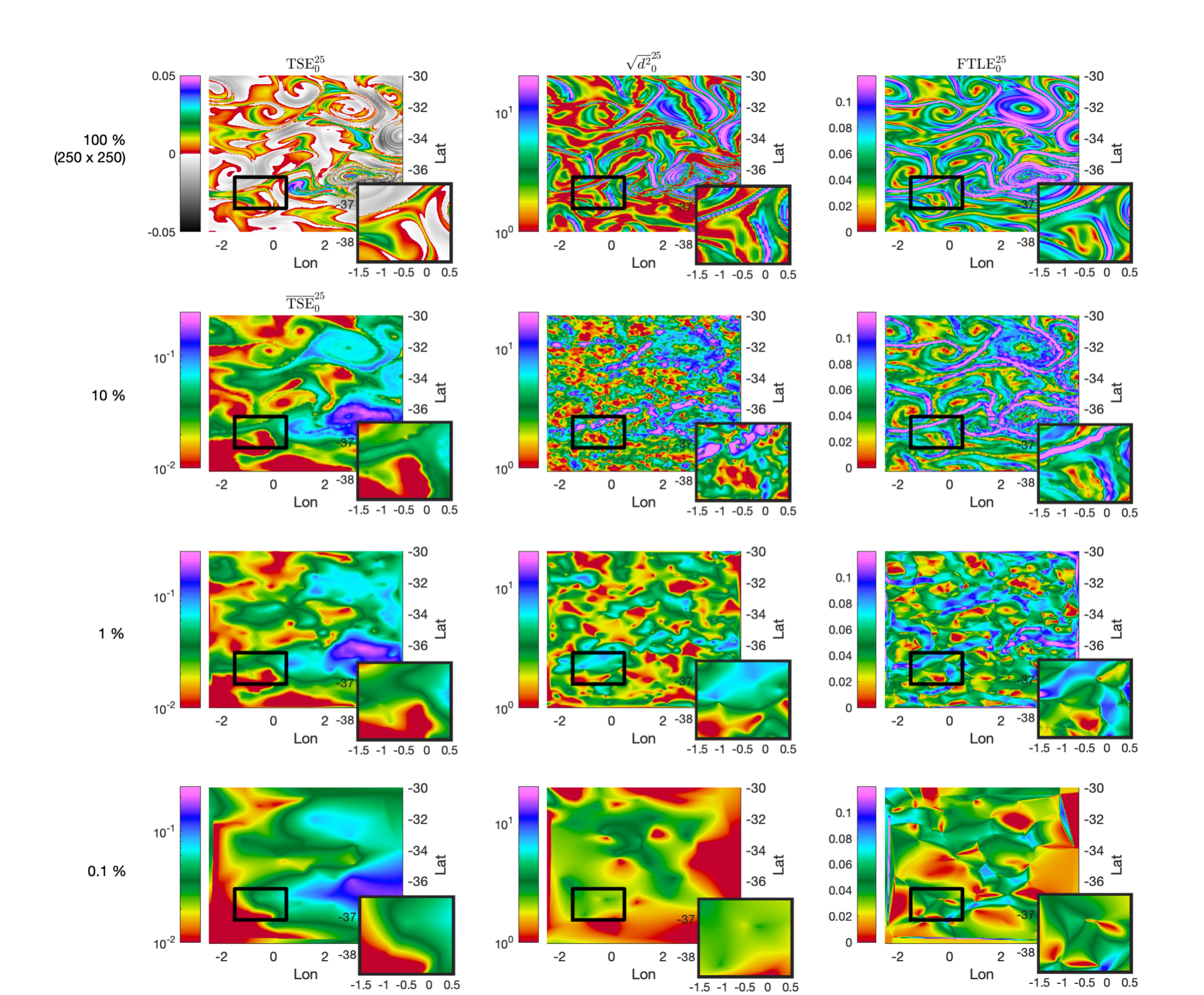} \caption{Stretching metric comparisons for AVISO ocean surface current fields.
The top row shows computations at full resolution with the lower rows
having progressively reduced resolution by randomly subsampling with
fewer trajectories. \label{fig:Stretching-metric-comparisons}}
\end{figure}

The first row of Fig. \ref{fig:Stretching-metric-comparisons} shows
the full-resolution computation of $\mathrm{TSE}_{0}^{25}(\boldsymbol{x}\mathbf{}_{0};0.2)$,
${d}^{2}(\boldsymbol{x}\mathbf{}_{i0},0,25)$ and $\mathrm{FTLE}_{0}^{25}(\boldsymbol{x}\mathbf{}_{0})$,
with an inset zoom of a front that separates two recirculation regions
in the flow. Successive rows gradually degrade from $10\%$ to $0.1\%$
of the full resolution. In these rows, we replace the stretching diagnostic
$\mathrm{TSE}_{0}^{25}(\boldsymbol{x}\mathbf{}_{0};0.2)$ with the
hyperbolicity strength diagnostic $\mathrm{\overline{TSE}}_{0}^{25}(\boldsymbol{x}\mathbf{}_{0};0.2)$,
which is a more robust single-trajectory indicator for sparse data.
The highlighted repelling LCS is evident in the top row as a ridge
in all three diagnostics. The quality of FTLE quickly degrades as
interpolations between final positions of particles oversimplify the
flow dynamics. The FTLE ridge of interest is still evident at the
$10\%$ subsampling, but it disappears below that resolution along
with many other previously distinguishable features. The relative
dispersion $d^{2}$ does a good job of approximating FTLE ridges at
full resolution, and provides a hint at some circulation patterns
at lower resolutions, but displays no clear separating feature reminiscent
of a ridge at any degree of subsampling.

As TSE values at a given location do not rely on any nearby neighbors,
their pointwise values remain unchanged and their features are more
resilient under subsampling in comparison to those of multi-trajectory
objective LCS metrics. Boundaries of multiple vortical features in
the flow, such as those shown in the inset, can be tracked through
multiple reductions in resolution as transitions between blue elliptic
features and the surrounding green turbulent regions. In fact, at
$0.1\%$ of the full resolution, there is still evidence of several
of the original major Lagrangian eddies in blue, as well as the front
in our inset.

Next we evaluate the ability of the $\overline{\mathrm{TRA}}$, as
defined in Theorem 3, to identify vortical features. We apply the
same method of random subsampling to compare $\mathrm{\overline{\mathrm{TRA}}}_{0}^{25}$
to the polar rotation angle, $\mathrm{PRA}_{0}^{25}$, whose features
(i.e., level curves) are objective in two dimensions \cite{farazmand16}.
In order to compute the flow map for the purposes of calculating PRA
from sparse data, we employ the same $C^{1}$ interpolant on final
trajectory positions that we used for computing $\mathrm{FTLE}_{0}^{25}$.
No objective, sparse-data-rotation-specific metric is available, which
prompts us to use $d^{2}$ instead in our comparison of rotation metrics.
We do expect $d^{2}$ to highlight elliptic LCSs as coherent Lagrangian
vortices, as observed first by \cite{provenzale99}.

Figure \ref{fig:Rotation-metric-comparisons} focuses on the same
time period and Agulhas region as Fig. \ref{fig:Stretching-metric-comparisons}.
Inset in each plot is a zoom on a region with a vortex previously
identified by exact mathematical methods as a black-hole-type elliptic
LCS (see e.g. \cite{haller13}). In the first row of plots, at full
resolution, all three metrics ($\mathrm{\overline{\mathrm{TRA}}}_{0}^{25}(\boldsymbol{x}\mathbf{}_{0};0.2)$,
${d}^{2}(\boldsymbol{x}\mathbf{}_{i0},0,25)$ and $\mathrm{PRA}_{0}^{25}(\boldsymbol{x}\mathbf{}_{0})$)
highlight this vortex, as well as several other elliptic LCSs in the
flow. The degree of detail provided by the objective, multi-trajectory
diagnostic, PRA, is not matched by the trajectory rotation metric
$\overline{\mathrm{TRA}}_{0}^{25}$, which nevertheless correctly
highlights the vortex.

The insensitivity of $\mathrm{\overline{\mathrm{TRA}}}_{0}^{25}(\boldsymbol{x}\mathbf{}_{0};0.2)$
to small details becomes an advantage at lower resolutions, as shown
by subsequent rows of Fig. \ref{fig:Rotation-metric-comparisons}.
Indeed, $\mathrm{\overline{\mathrm{TRA}}}_{0}^{25}(\boldsymbol{x}\mathbf{}_{0};0.2)$
retains much of its structure with progressive random subsamplings.
As a result, it is still possible to recognize three of its maxima
on the right-hand-side, corresponding to multiple vortices, even at
the lowest resolution. Some strong rotational features can be tracked
below $10\%$ resolution in the PRA fields, but the elliptic LCS quickly
becomes indistinguishable. Similarly, distinguishing coherent structures
in the $\sqrt{d^{2}}$ plots becomes unfeasible quickly under progressive
subsampling,

For a comparison with a non-objective diagnostic suggested for sparse
geophysical data, we consider the mean of extrema extents ($\mathrm{MEEX}$)
diagnostic proposed by \cite{mundel14}. As we noted in the introduction,
this diagnostic has no known relation to material stretching or rotation
but its evaluation is compellingly simple. Indeed, for a particle
trajectory initiated at $\boldsymbol{x}_{0}\in U$, the $\mathrm{MEEX}_{t_{0}}^{t_{f}}(\boldsymbol{x}_{0})$
field is simply defined as

\begin{equation}
\mathrm{MEEX}_{t_{0}}^{t_{N}}(\boldsymbol{x}_{0})=\frac{\min_{t\in[t_{0},t_{N}]}\phi(\boldsymbol{x}(t))+\max_{t\in[t_{0},t_{N}]}\phi(\boldsymbol{x}(t))}{2},
\end{equation}
where $[t_{0},t_{N}]$ is the observation interval and $\phi(\boldsymbol{x})$
is a scalar-valued observable. The fluid particle zonal (longitudinal)
or meridional (latitudinal) extent (measured in km) was suggested
in \cite{mundel14} as a suitable $\phi$ for sparse ocean drifter
data. 

\begin{figure}[t]
\centering \includegraphics[scale=0.45]{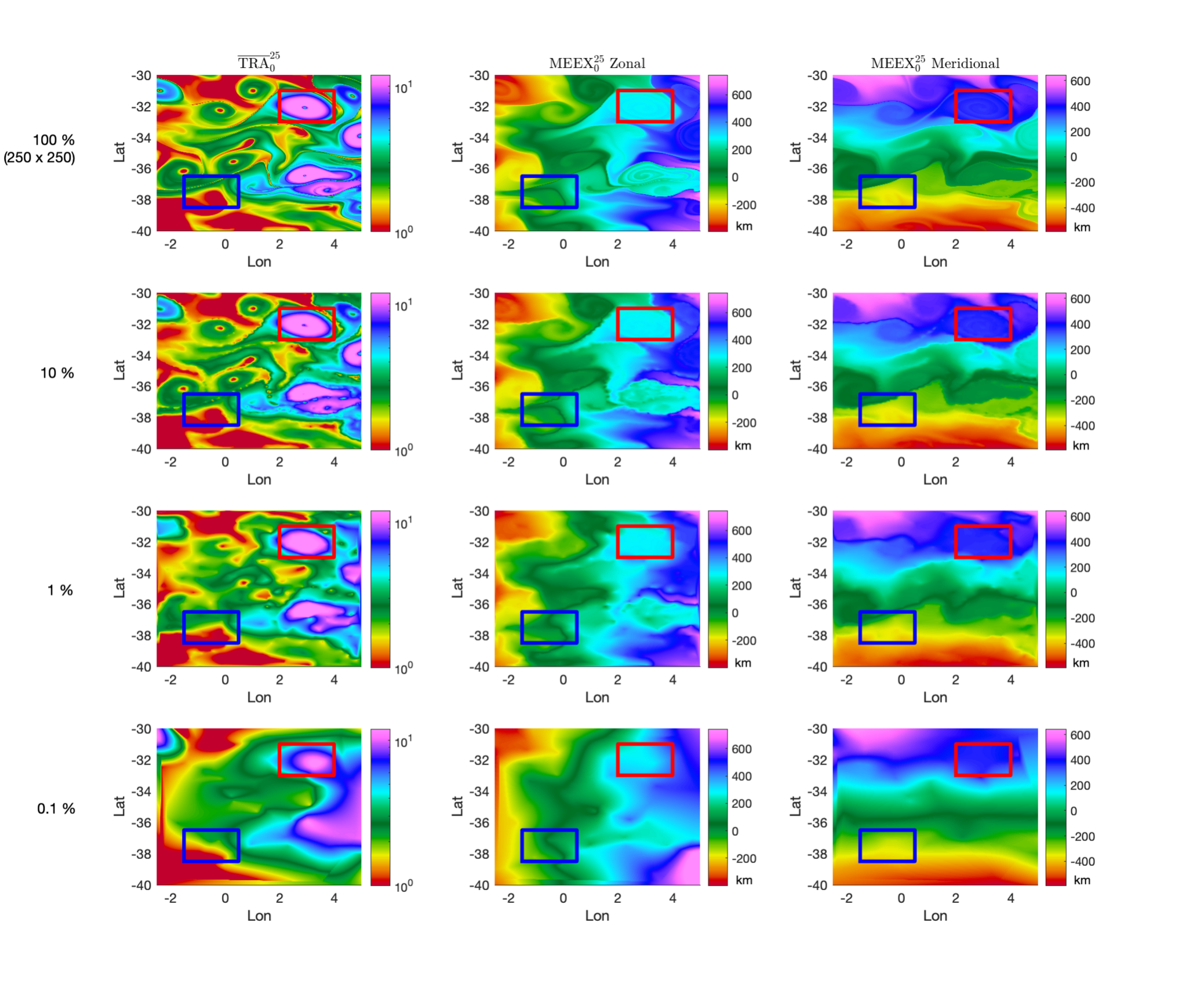} \caption{Comparison of $\mathrm{\overline{\mathrm{TRA}}}_{0}^{25}(\boldsymbol{x}\mathbf{}_{0};0.2)$
with the $\mathrm{MEEX}_{0}^{25}(\boldsymbol{x}_{0})$ diagnostic
fields proposed in \cite{mundel14}, calculated in the zonal and meridional
directions for the Agulhas leakage. Blue boxes indicate the location
of frontal features from Fig. \ref{fig:Stretching-metric-comparisons}
and red boxes highlight the vortex from Fig. \ref{fig:Rotation-metric-comparisons}.}
\label{fig:MEEX} 
\end{figure}

The left column of Fig. \ref{fig:MEEX} shows $\mathrm{\overline{\mathrm{TRA}}}_{0}^{25}(\boldsymbol{x}\mathbf{}_{0};0.2)$
calculated over the same spatial-temporal domain as Fig. \ref{fig:Rotation-metric-comparisons},
with the same degrees of data degradation. The middle column shows
$\mathrm{MEEX}_{0}^{25}(\boldsymbol{x}_{0})$ with $\phi$ equal to
the zonal distance of fluid particles from the prime meridian, and
the right column shows $\mathrm{MEEX}_{0}^{25}(\boldsymbol{x}_{0})$
with $\phi$ equal to the meridional distance from the $-35^{\circ}$
line of latitude. As expected from \cite{mundel14}, the middle column
corresponds with the best performing version of MEEX for full resolution
current data in the Agulhas region. The blue boxes in each subplot
correspond with the front highlighted in the previous TSE comparison
(Fig. \ref{fig:Stretching-metric-comparisons}) and the red boxes
correspond with the vortex in Fig. \ref{fig:Rotation-metric-comparisons}. 

Overall, the MEEX diagnostics shows correlations with the Lagrangian
features revealed by objective diagnostics in Fig. \ref{fig:Rotation-metric-comparisons}
and confirmed by particle advection in various earlier studies of
the same data set (\cite{beronvera13,haller16b,hadjighasem16,aksamit20}).
This is to be expected, as any generic observable evaluated on trajectory
positions will be influenced by LCSs, as demonstrated by \cite{hadjighasem17}
on several examples. The details of the MEEX patterns and their closeness
to LCSs, however, depend on the choice of $\phi(\boldsymbol{x})$. 

Under subsampling of the data, the hyperbolic and elliptic structures
persist in the $\mathrm{\overline{\mathrm{TRA}}}_{0}^{25}(\boldsymbol{x}\mathbf{}_{0};0.2)$
field and several patterns in $\mathrm{MEEX}_{0}^{25}(\boldsymbol{x}_{0})$
are also resilient. However, while patterns suggesting elliptic LCSs
are initially visible in the full-resolution zonal and meridional
$\mathrm{MEEX}_{0}^{25}$ fields, they tend to disappear in the degraded
data. The mixing behavior across the front in the blue box is also
difficult to interpret in the $\mathrm{MEEX}_{0}^{25}$ fields, with
no indication of a major transport barrier at or below $1\%$. Combining
interpretations of the two $\mathrm{MEEX}_{0}^{25}$ fields presents
a further challenge due to the seemingly contradictory patterns that
arise from the different choices of spatial directions for the observable
$\phi(\boldsymbol{x})$.

\subsection{Unsteady ABC Flow\label{subsec:Unsteady ABC}}

As a second example, we consider the three-dimensional velocity field

\begin{equation}
\mathbf{v}(\boldsymbol{x},t)=e^{-\nu t}\begin{pmatrix}A\sin x_{3}+C\cos x_{2}\\
B\sin x_{1}+A\cos x_{3}\\
C\sin x_{2}+B\cos x_{1}
\end{pmatrix},\label{eq:unsteady ABC velocity field}
\end{equation}
a viscous, unsteady version of the steady ABC flow, defined on the
triply periodic spatial domain $U=[0,2\pi]^{3}.$ As in ref. \cite{farazmand16},
we use the parameter values $A=1$, $B=\sqrt{2/3}$, $C=\sqrt{1/3}$,
with the added viscosity value $\nu=0.01$.

The unsteadiness of the flow necessitates the use of Theorem 3. By
the triple periodicity of $\mathbf{v}(\boldsymbol{x},t)$ on $U$,
we also have $\bar{\boldsymbol{\omega}}(t)\equiv\mathbf{0}$ and hence
condition (\textbf{A2}) is also satisfied on $U$. We conclude that
both TSE metrics are objective on the extended phase space and the
$\overline{\mathrm{TRA}}(\boldsymbol{x}\mathbf{}_{0};\mathrm{v}_{0})$
metric is quasi-objective on the extended phase space in the frame
of (\ref{eq:unsteady ABC velocity field}). As the ABC flow is a dimensionless
set of equations, we use $\mathrm{v}_{0}=1$.

\begin{figure}[t]
\centering \includegraphics[scale=0.45]{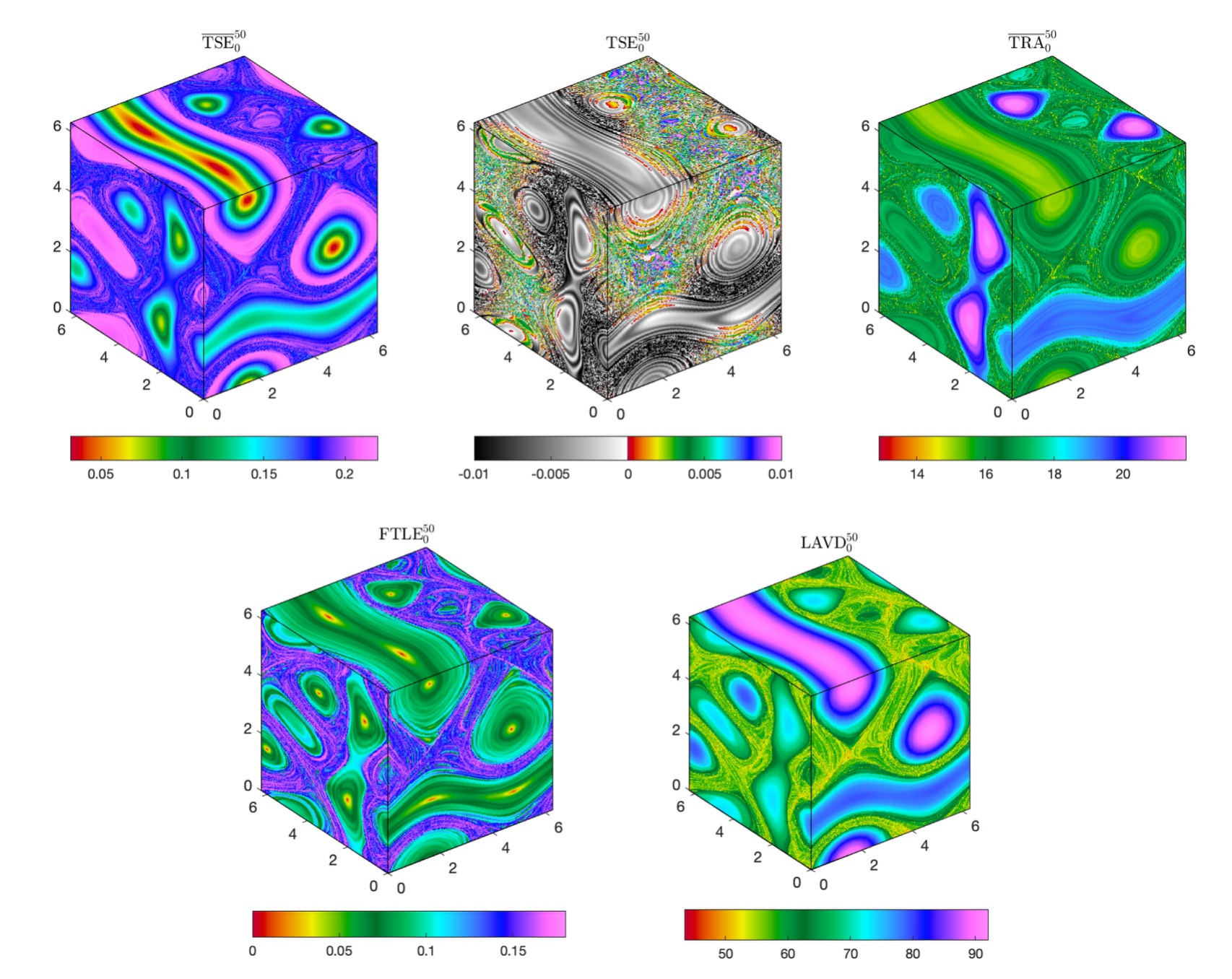} \caption{Elliptic and hyperbolic LCSs in the unsteady ABC flow \eqref{eq:unsteady ABC velocity field}.
The plots compare quasi-objective, single-trajectory metrics (\textbf{$\mathrm{TSE_{0}^{50}(\boldsymbol{x}\mathbf{}_{0};1)}$
}and\textbf{ $\mathrm{\mathrm{\overline{\mathrm{TRA}}{}_{0}^{50}}}(\boldsymbol{x}\mathbf{}_{0};1)$})
with objective LCS metrics (\textbf{$\mathrm{FTLE}_{0}^{50}(\boldsymbol{x}\mathbf{}_{0})$
}and\textbf{ $\mathrm{LAVD}_{0}^{50}(\boldsymbol{x}\mathbf{}_{0})$})
that require multiple neighboring trajectories or detailed knowledge
of the velocity field. \label{fig:ABC flow high res}}
\end{figure}

In Fig. \ref{fig:ABC flow high res}, we show plots of the $\mathrm{TSE}_{0}^{50}(\boldsymbol{x}\mathbf{}_{0};1)$,
$\mathrm{\overline{TSE}}_{0}^{50}(\boldsymbol{x}\mathbf{}_{0};1)$
and $\mathrm{\overline{\mathrm{TRA}}_{0}^{50}(\boldsymbol{x}\mathbf{}_{0};1)}$
metrics computed from Theorem 3 over trajectories launched from an
initial grid of $200\times200\times200$ points, integrated from time
$t_{0}=0$ to $t_{N}=50$. Values of the FTLE and LAVD are also included
for comparison. (We did not compute PRA for this example as levels
sets of the PRA are not objective in three dimensions.) We conclude
that the quasi-objective, single-trajectory diagnostics ($\mathrm{TSE_{0}^{50}}$,
$\mathrm{\overline{TSE}_{0}^{50}}$, $\mathrm{\overline{\mathrm{TRA}}_{0}^{50}}$)
perform just as well as the multi-trajectory, objective metric, the
$\mathrm{FTLE}_{0}^{50}$, and the single-trajectory, objective, but
velocity-field-reliant metric, the $\mathrm{LAVD}_{0}^{50}$.

To evaluate the performance of the quasi-objective metrics under a
random degradation of the initial grid, we employ the same random
subsampling method as in \ref{subsec:2D ocean surface data}. Even
though level sets of the PRA are not objective in three-dimensions,
we include PRA here in our comparison. This is because there is no
systematic or widely accepted procedure for accurately approximating
vorticity from sparse Lagrangian trajectories for LAVD calculations.

\begin{figure}[t]
\centering \includegraphics[scale=0.4]{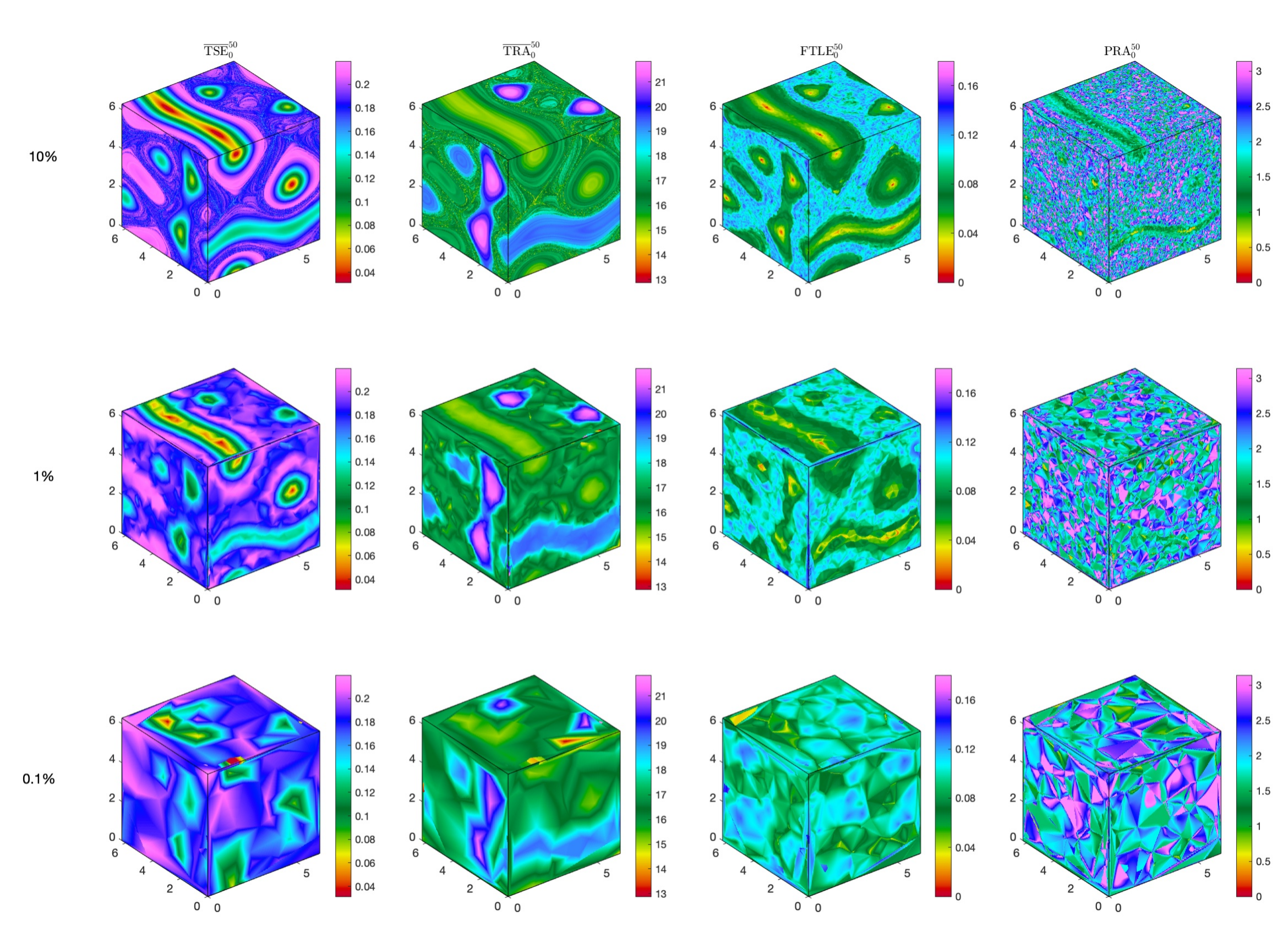} \caption{Elliptic and hyperbolic LCS in the randomly subsampled unsteady ABC
flow \eqref{eq:unsteady ABC velocity field}. The plots compare quasi-objective
single-trajectory metrics (\textbf{$\mathrm{\overline{TSE}{}_{0}^{50}}(\boldsymbol{x}\mathbf{}_{0};1)$
}and\textbf{ $\mathrm{\overline{TRA}_{0}^{50}}(\boldsymbol{x}\mathbf{}_{0};1)$})
with LCS metrics (\textbf{$\mathrm{FTLE}_{0}^{50}(\boldsymbol{x}\mathbf{}_{0})$
}and\textbf{ $\mathrm{PRA}_{0}^{50}(\boldsymbol{x}\mathbf{}_{0})$}),
whose computation requires multiple neighboring trajectories. The
initial condition grid for trajectories is randomized and its density
is gradually decreased to $0.1\%$ of its initial value. \label{fig:ABC flow sparsified}}
\end{figure}

Figure \ref{fig:ABC flow sparsified} shows that $\mathrm{\overline{TSE}_{0}^{50}}$
and \textbf{$\mathrm{\mathrm{\overline{\mathrm{TRA}}{}_{0}^{50}}}$}
continue to provide information on dominant flow features even as
trajectory data becomes substantially sparse. Specifically, as we
keep the bounds on the colormaps constant, the separation between
rotational regions remains clear from $\mathrm{\overline{TSE}_{0}^{50}}$
and \textbf{$\mathrm{\mathrm{\mathrm{\overline{\mathrm{TRA}}{}_{0}^{50}}}}$}
when compared with \textbf{$\mathrm{FTLE}_{0}^{50}$ }and\textbf{
$\mathrm{PRA}_{0}^{50}$}, down to a random selection of trajectories
consisting of $0.1\%$ of the original computation grid. We conclude
that even in three-dimensional unsteady flows, it is possible to observe
a consistent range of stretching and rotation rates with single-trajectory,
quasi-objective metrics from very sparse and randomly positioned data.
Furthermore, interpolating between randomly oriented $\mathrm{TSE}$
and $\mathrm{TRA}$ metrics provides a diagnostic that highlights
time-varying coherent structures to an extent that is unreachable
by prior LCS methods.

\subsection{Momentum barriers in the unsteady ABC flow\label{subsec:Momentum-barriers-in ABC flow}}

As our third example, we consider the linear-momentum barrier field
equations, as defined in \cite{haller20}, for the unsteady ABC flow
(\ref{eq:unsteady ABC velocity field}). As shown in \cite{haller20},
at a time $t$, the instantaneous barriers to momentum transport are
invariant manifolds of the autonomous dynamical system

\begin{equation}
\boldsymbol{x}'=\mathbf{w}(\mathbf{x};t)=\nu\rho\Delta\boldsymbol{\mathbf{v}}(\boldsymbol{x},t),\label{eq:barrier field}
\end{equation}
with prime denoting differentiation with respect to a fictitious time
(barrier time) parametrizing the curves forming the barrier surface.
At time $t=0$, eq. \eqref{eq:barrier field} simplifies to $\boldsymbol{x}'=-\nu\rho\mathbf{v}(\boldsymbol{x},0)$
in the case of the unsteady ABC flow defined in (\ref{eq:unsteady ABC velocity field}).
This autonomous dynamical system provides a steady flow field on which
we can compute coherent structure diagnostics to highlight instantaneous
barriers to the transport of momentum in the underlying fluid flow.
This allows us to compare the steady versions of the (now objective)
single trajectory metrics $\mathrm{\overline{TSE}{}_{0}^{50}}(\boldsymbol{x}\mathbf{}_{0})$,
$\mathrm{TSE}{}_{0}^{50}(\boldsymbol{x}\mathbf{}_{0})$, and $\mathrm{\overline{TRA}{}_{0}^{50}(\boldsymbol{x}\mathbf{}_{0})}$,
as defined in Theorems 1 and 2, with the objective LCS metrics $\mathrm{FTLE}_{0}^{50}(\boldsymbol{x}\mathbf{}_{0})$
and $\mathrm{LAVD}_{0}^{50}(\boldsymbol{x}\mathbf{}_{0})$. As these
active barrier vector fields are objective \cite{haller20}, we can
now also include a comparison with objective $\mathrm{TRA}{}_{0}^{50}(\boldsymbol{x}\mathbf{}_{0})$,
as defined in (\ref{eq:objective alpha}) (see Remarks 3 and 1 after
Theorems 1 and 2, respectively, on the objectivity of these metrics
for $\mathbf{w}(\mathbf{x};t)$). In our comparison, we will use the
same parameters as in the previous section for the flow (\ref{eq:unsteady ABC velocity field}).

\begin{figure}[t]
\centering \includegraphics[scale=0.43]{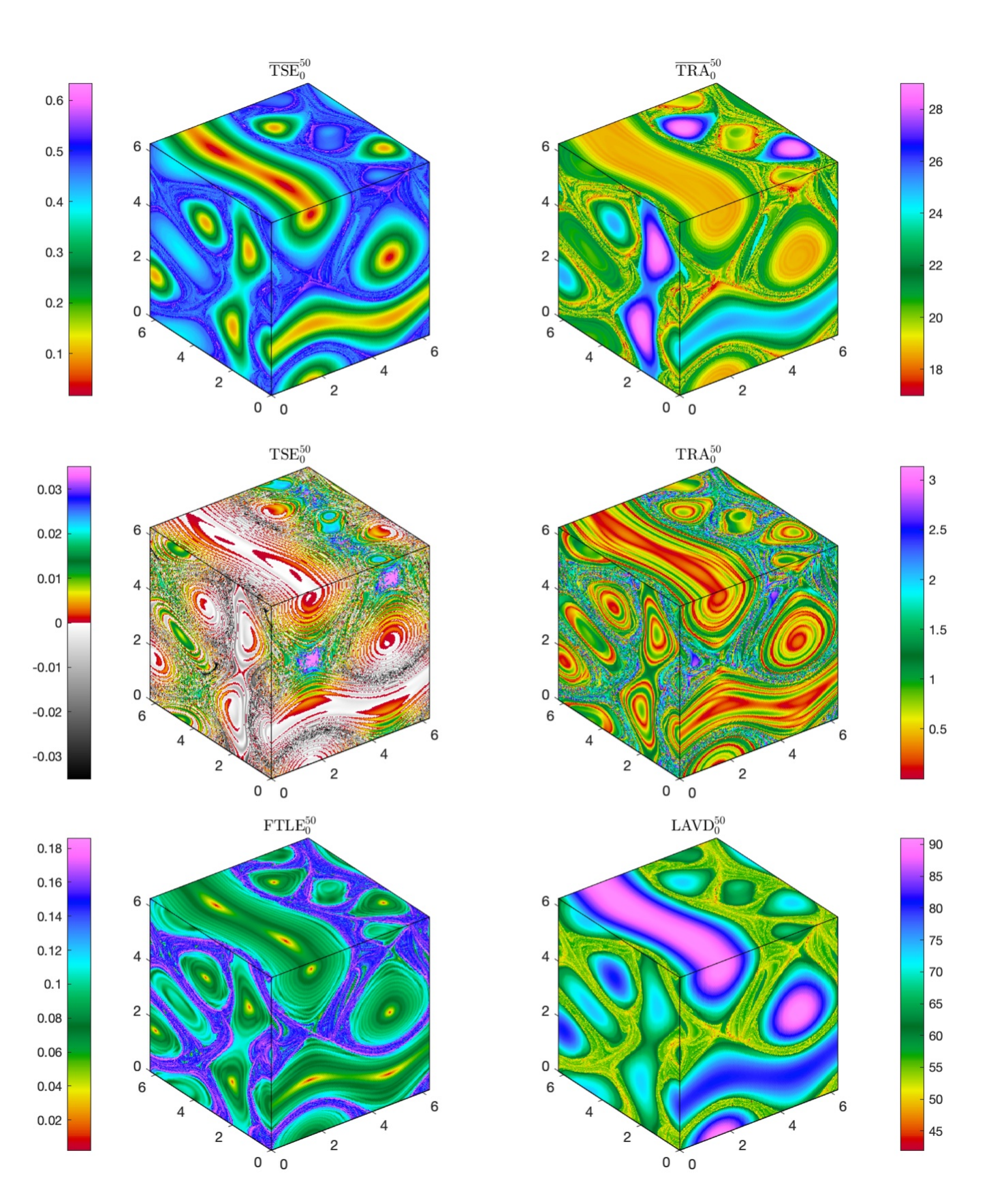} \caption{Elliptic and hyperbolic barriers of instantaneous linear momentum
transport in the unsteady ABC flow \eqref{eq:unsteady ABC velocity field}.
The plots compare the objective single-trajectory metrics (\textbf{$\mathrm{\overline{TSE}{}_{0}^{50}(\boldsymbol{x}\mathbf{}_{0})}$
}and\textbf{ $\mathrm{\overline{TRA}_{0}^{50}(\boldsymbol{x}\mathbf{}_{0})}$};
see remarks 3. and 1. after Theorems 1 and 2, respectively) with the
objective LCS metrics \textbf{$\mathrm{FTLE}_{0}^{50}(\boldsymbol{x}\mathbf{}_{0})$
}and\textbf{ $\mathrm{PRA}_{0}^{50}(\boldsymbol{x}\mathbf{}_{0})$},
whose computation requires multiple neighboring trajectories or detailed
knowledge of the velocity field. \label{fig:ABC barriers high res}}
\end{figure}

\begin{figure}[t]
\centering \includegraphics[scale=0.4]{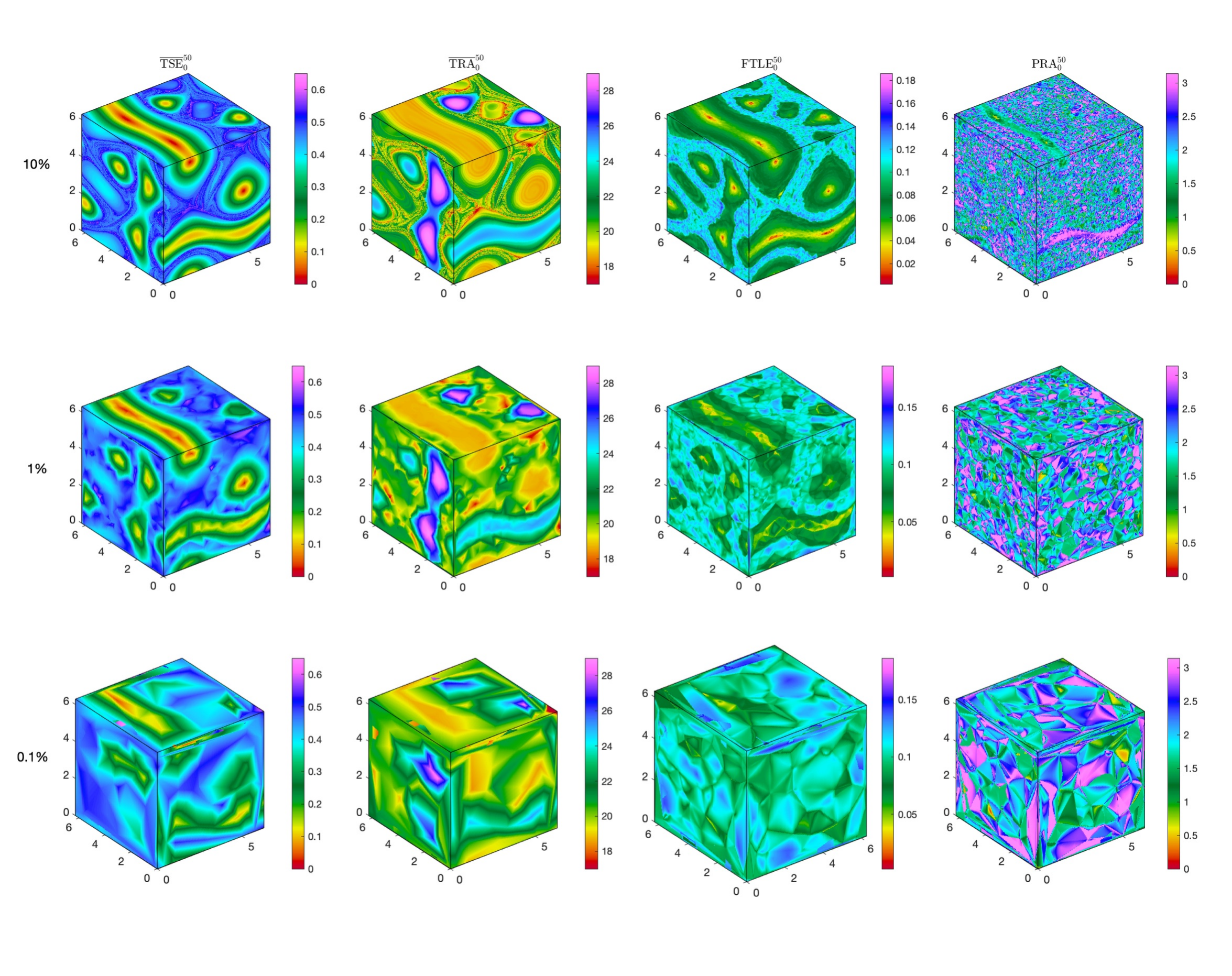} \caption{Same as Fig. \ref{fig:ABC barriers high res} but under subsampling
of the unsteady ABC flow \eqref{eq:unsteady ABC velocity field}.
The initial condition grid for trajectories is randomized and its
density is gradually decreased to $0.1\%$ of its initial value. \label{fig:ABC barriers sparsified}}
\end{figure}

Figure \ref{fig:ABC barriers high res} shows the ability of the single-trajectory
metrics to accurately represent the elliptic and hyperbolic structures
in the active barrier vector field \eqref{eq:barrier field} for trajectories
with initial positions on the same rectangular grid in the domain
$U$ as in Section \ref{subsec:Unsteady ABC}. When compared with
$\mathrm{FTLE}_{0}^{50}$, we find that the steady-version of $\mathrm{TSE_{0}^{50}}$
provides a great degree of detail, highlighting hyperbolic regions
of the flow as before. In addition, $\mathrm{TRA_{0}^{50}}$ is able
to provide a comparable amount of detail for elliptic features. As
we degrade the resolution with the same random subsampling of initial
positions, Fig. \ref{fig:ABC barriers sparsified} shows that $\overline{\mathrm{TSE}}_{0}^{50}$
and $\overline{\mathrm{TRA}}_{0}^{50}$ are again robust and able
to highlight hyperbolic and elliptic transport barriers down to $0.1\%$
of the original resolution. Calculations of $\mathrm{FTLE}_{0}^{50}$
and $\mathrm{PRA}_{0}^{50}$ from sparse trajectory data on the steady
barrier field perform comparably to the unsteady flow example, showing
a quick deterioration in quality and interpretability.

\section{Conclusions}

Coherent structures, such as fronts, jets and vortices, are best viewed
as material objects. Indeed, these structures are most reliably identified
in experiments and observations from their signatures in material
tracer fields. By definition, such material coherent structures are
indifferent to the observer and hence can only be self-consistently
identified from objective quantities. In contrast, flows of practical
relevance are mostly known from trajectory data, which is inherently
non-objective. Specifically, all features of trajectories (including
their length, curvature, velocity and acceleration) are dependent
on the observer. The question, therefore, naturally arises: How can
one still use trajectory data to extract reliable information from
material coherent structures in a flow?

We have addressed this question here by introducing the notion of
quasi-objectivity. We call a scalar-, vector- or tensor-field quasi-objective
under a condition if the field approximates another, objective field
in all frames in which that condition is satisfied. We have derived
quasi-objective measures of trajectory stretching ($\mathrm{TSE}$
and $\mathrm{\overline{TSE}}$) that approximate objective material
stretching exponents in frames where the flow is steady, ie., satisfies
condition (\textbf{A1}). We have also derived quasi-objective measures
of trajectory rotation ($\mathrm{TRA}$ and $\mathrm{\overline{TRA}}$)
that approximate objective material rotation under conditions (\textbf{A1})
and (\textbf{A2}). The latter condition requires trajectory accelerations
to dominate the angular acceleration induced by the spatial mean vorticity.
Viewing general unsteady flows as steady flows in their extended phase
space (space-time), we have also obtained objective TSE metrics and
quasi-objective TRA metrics under assumption (\textbf{A2}) on the
extended phase space. These extended metrics track the stretching
and rotation of tangent vectors to fluid trajectories in space-time.

We have tested these quasi-objective diagnostics on two-dimensional
mesoscale ocean surface velocity data (AVISO) obtained from satellite
altimetry. Condition (\textbf{A2}) is clearly satisfied on regions
large enough to contain several mesoscale eddies. Accordingly, we
have found the quasi-objective metrics to perform very well. As a
further test case, we have considered an unsteady version of the classic
ABC flow \cite{haller20} which is an exact solution of the three-dimensional
unsteady Navier--Stokes equation. This flow satisfies condition (\textbf{A2})
on its triply periodic domain of definition. We have also considered
the steady version, which satisfies assumption (\textbf{A1}).

On trajectories generated by these example velocity fields from sufficiently
dense grids, the trajectory stretching exponent ($\mathrm{TSE}$)
faithfully reproduced features of the FTLE field. This is notable
because $\mathrm{TSE}$ is a single-trajectory-based diagnostic whose
pointwise value is independent from the number and location of other
trajectories in the data set. For sparse and irregularly spaced trajectory
data, the trajectory hyperbolicity strength ($\mathrm{\overline{TSE}}$)
returned distinct indications of coherent structures even when FTLE
and relative dispersion computations gave no meaningful results. Similarly,
we have found that on well-resolved trajectory data from objective
vector fields, the trajectory rotation angle ($\mathrm{TRA}$) accurately
reproduced details of the PRA field, even though $\mathrm{TRA}$ is
a single-trajectory-based diagnostic. For sparse and irregularly spaced
trajectory data, the cancellation-free total rotation angle ($\mathrm{\overline{TRA}}$)
gave clear indication of coherent structures even when PRA computations
were no longer feasible. All this suggest great potential for $\mathrm{\overline{TSE}}$
and \textbf{$\mathrm{\overline{TRA}}$} in visualizing coherent structures
in 3D particle tracking velocimetry experiments, in which tracer concentrations
are generally low (see, e.g., \cite{hoyer05,rosi14}). In contrast,
$\mathrm{TSE}$ and (for objective vector fields, such as the barrier
equations in \cite{haller20} ) the $\mathrm{TRA}$, are useful in
reducing computational costs relative to multi-trajectory LCS diagnostics
when trajectories from a well-resolved initial grid are available.

Further applications of the quasi-objective diagnostic involve the
detection of barriers to the transport of dynamically active vector
fields, such as the momentum and the vorticity \cite{haller20}. These
barriers are invariant manifolds of appropriate steady, volume-preserving
and objective vector fields, the barrier vector fields. The computation
of these barrier vector fields requires two spatial differentiations
of the underlying velocity field and hence their analysis via FTLE
and PRA is numerically challenging, given that the latter two methods
involve further differentiation. In this setting, condition (\textbf{A1})
is always satisfied and (\textbf{A2}) can be waived without any loss
of generality, as we have pointed out. Consequently, the use of $\mathrm{TSE}$
and $\mathrm{TRA}$ promises to bring a major improvement in computational
efficiency for the detection of active transport barriers in three-dimensional
unsteady flows.

\vskip 1 true cm

\textbf{Acknowledgment}

The authors acknowledge financial support from Priority Program SPP
1881 (Turbulent Superstructures) of the German National Science Foundation
(DFG). \vskip 2 true cm \textbf{Data Availability}

The AVISO geostrophic current velocity product used in this study,
``Global Ocean Gridded L4 Sea Surface Heights and Derived Variables
Reprocessed,'{}' is freely available and is hosted by the Copernicus
Marine Environment Monitoring Service (http://marine.copernicus.eu).

\end{document}


\title{Erratum:''Quasi-objective coherent structure diagnostics \\
 from single trajectories'' {[}Chaos 31, 043131 (2021){]}}
\author{George Haller\thanks{Email:georgehaller@ethz.ch}, Nikolas Aksamit
and Alex P. Encinas-Bartos\\
 Institute for Mechanical Systems\\
 ETH Zürich, 8092 Zürich, Switzerland\\
 }

\maketitle
This erratum corrects a mistake in our previously published paper
\cite{haller21}. Theorem 3 is incorrect as stated because the extended
Euclidean frame changes of the form (32) are nonlinear transformations
of the extended phases space. Their Jacobians generally does not preserve
lengths and angles, and hence the scalar fields $\mathrm{TSE}$, $\overline{\mathrm{TSE}}$
and $\mathrm{\overline{\mathrm{TRA}}}$ are not invariant under these
transformations.

Theorem 3 can be corrected by modifying assumption (A1) for unsteady
flows as\\

\begin{description}
\item [{(A3)~~~~~~~~~~~~~~~~~~~~~~~~~~~~~~~~~~~~~~~~~~~~~~~~$\left|\partial_{t}\mathbf{v}(\mathbf{x}(t),t)\right|\ll\left|\mathbf{a}(t)\right|$,}] 
\end{description}
where $\mathbf{a}(t):=\ddot{\mathbf{x}}(t)$ denoted the Lagrangian
acceleration. This reflects the assumption that Lagrangian time scales
dominate Eulerian time scales in geophysical flows, which originally
motivated our study. On specific flow domains, (A3) can be a priori
verified from characteristic velocity-, length- and time-scales (see,
e.g, Ref. \cite{pedlosky}, p. 88). Assumption (A3) also eliminates
the need to use the extended phase space in our arguments and enables
us to set $\mathrm{v}_{0}=0$ in our formulas. Figure 1. shows that
(A3) is satisfied on average for both the AVISO data set and the unsteady
ABC flow example we considered in \cite{haller21}.
\begin{figure}
\centering \includegraphics[width=1\textwidth]{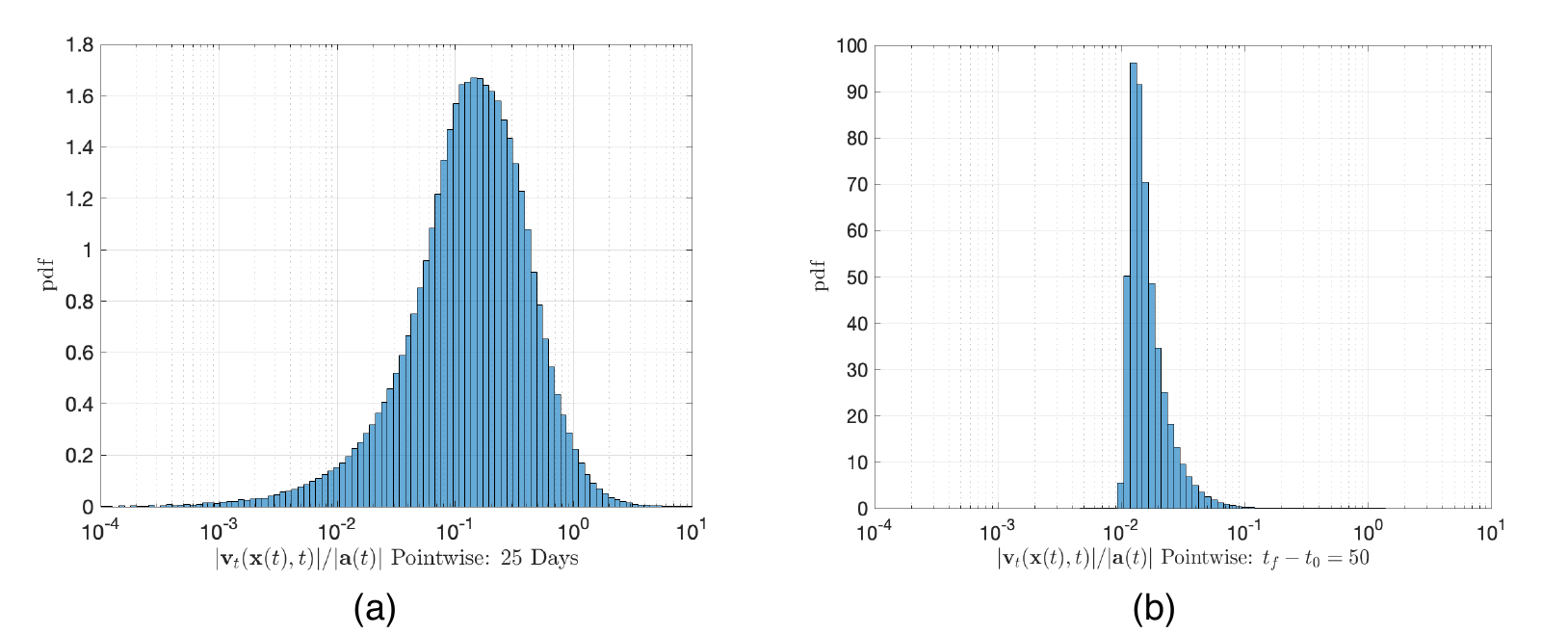}\caption{Verification of assumption (A3) for our two examples in \cite{haller21}:
Probability density function (pdf) of the values of $\left|\partial_{t}\mathbf{v}(\mathbf{x}(t),t)\right|/\left|\mathbf{a}(t)\right|$
for assumption (A3) in (a) the AVISO data set example (b) the unsteady
ABC flow example. The pdf is measured in probability per unit increment.}
\end{figure}

\textcolor{black}{The evolution of the Lagrangian velocity $\boldsymbol{v}(t):=\mathbf{v}(\mathbf{x}(t),t)$
is nearly material in frames satisfying (A3). Additionally, such frames
must be related to each other  via Euclidean coordinate transformations
$\mathbf{x}=\mathbf{Q}(t)\mathbf{y}+\mathbf{b}(t)$ with $\left|\dot{\mathbf{Q}}\right|,\left|\dot{\mathbf{b}}\right|,\left|\ddot{\mathbf{Q}}\right|,\left|\ddot{\mathbf{b}}\right|\ll1$,
i.e., via slowly varying (geophysical) frame changes, in order for
the $\mathrm{TSE}$, $\overline{\mathrm{TSE}}$ and $\mathrm{\overline{\mathrm{TRA}}}$
to return approximately the same values in the new frames. More specifically,
our revised definition of quasi-objectivity for a scalar field is
that it has to approximate the same objective quantity in all frames
related to each other by slowly varying Euclidean transformation. }

\textcolor{black}{In any frame satisfying (A3), therefore, the single-trajectory
diagnostics $\lambda{}_{t_{0}}^{t_{N}}(\mathbf{\boldsymbol{x}}_{0},\boldsymbol{v}(t_{0}))$
and $\bar{\lambda}{}_{t_{0}}^{t_{N}}(\mathbf{\boldsymbol{x}}_{0},\boldsymbol{v}(t_{0}))$
will closely approximate the corresponding objective measures (averaged
stretching exponent and averaged hyperbolicity strength) of material
elements initially aligned with $\boldsymbol{v}(t_{0})$. This implies
the quasi-objectivity of these single-trajectory diagnostics by the
above revision of the definition of Ref. \cite{haller21}. A similar
statement holds for the rotation diagnostic $\bar{\alpha}_{t_{0}}^{t_{N}}(\mathbf{\boldsymbol{x}}_{0},\boldsymbol{v}(t_{0}))$
under assumptions (A2)-(A3). In summary, the corrected Theorem 3 reads
as follows:}
\begin{description}
\item [{Theorem~3.}]~

(i) Under assumption \textbf{(A3)}, the trajectory stretching exponents\emph{
}(TSEs), defined as 
\[
\mathrm{TSE}_{t_{0}}^{t_{N}}(\boldsymbol{x}\mathbf{}_{0})=\frac{1}{t_{N}-t_{0}}\log\frac{\left|\dot{\mathbf{x}}(t_{N})\right|}{\left|\dot{\mathbf{x}}(t_{0})\right|},\qquad\overline{\mathrm{TSE}}_{t_{0}}^{t_{N}}(\boldsymbol{x}\mathbf{}_{0})=\frac{1}{t_{N}-t_{0}}\sum_{i=0}^{N-1}\left|\log\frac{\left|\dot{\mathbf{x}}(t_{i+1})\right|}{\left|\dot{\mathbf{x}}(t_{i})\right|}\right|,
\]
are quasi-objective measures of trajectory stretching and hyperbolicity
strength.

(ii) Under assumptions (\textbf{A2})-\textbf{(A3)}, the trajectory
angular velocity\emph{ }($\mathrm{\overline{\mathrm{TRA}}}$), defined
as
\begin{align*}
\mathrm{\overline{\mathrm{TRA}}}_{t_{0}}^{t_{N}}(\boldsymbol{x}\mathbf{}_{0}) & =\frac{1}{t_{N}-t_{0}}\sum_{i=0}^{N-1}\cos^{-1}\frac{\left\langle \dot{\mathbf{\boldsymbol{x}}}(t_{i}),\dot{\mathbf{\boldsymbol{x}}}(t_{i+1})\right\rangle }{\left|\dot{\mathbf{\boldsymbol{x}}}(t_{i})\right|\left|\dot{\mathbf{\boldsymbol{x}}}(t_{i+1})\right|},
\end{align*}
is a quasi-objective measure of total trajectory rotation.
\end{description}
\begin{singlespace}
Figures 2-5 below are revised version of Figs. 1,2,4 and 5 of Ref.
\cite{haller21}, showing that our main conclusions remain valid under
the corrected implementation of Theorem 3. Figures 6 and 7 need no
revision, because their computations were carried out on an objective
vector field for which assumptions (A2) and (A3) are not required.
\end{singlespace}

\begin{figure}[t]
\centering \includegraphics[width=1\textwidth]{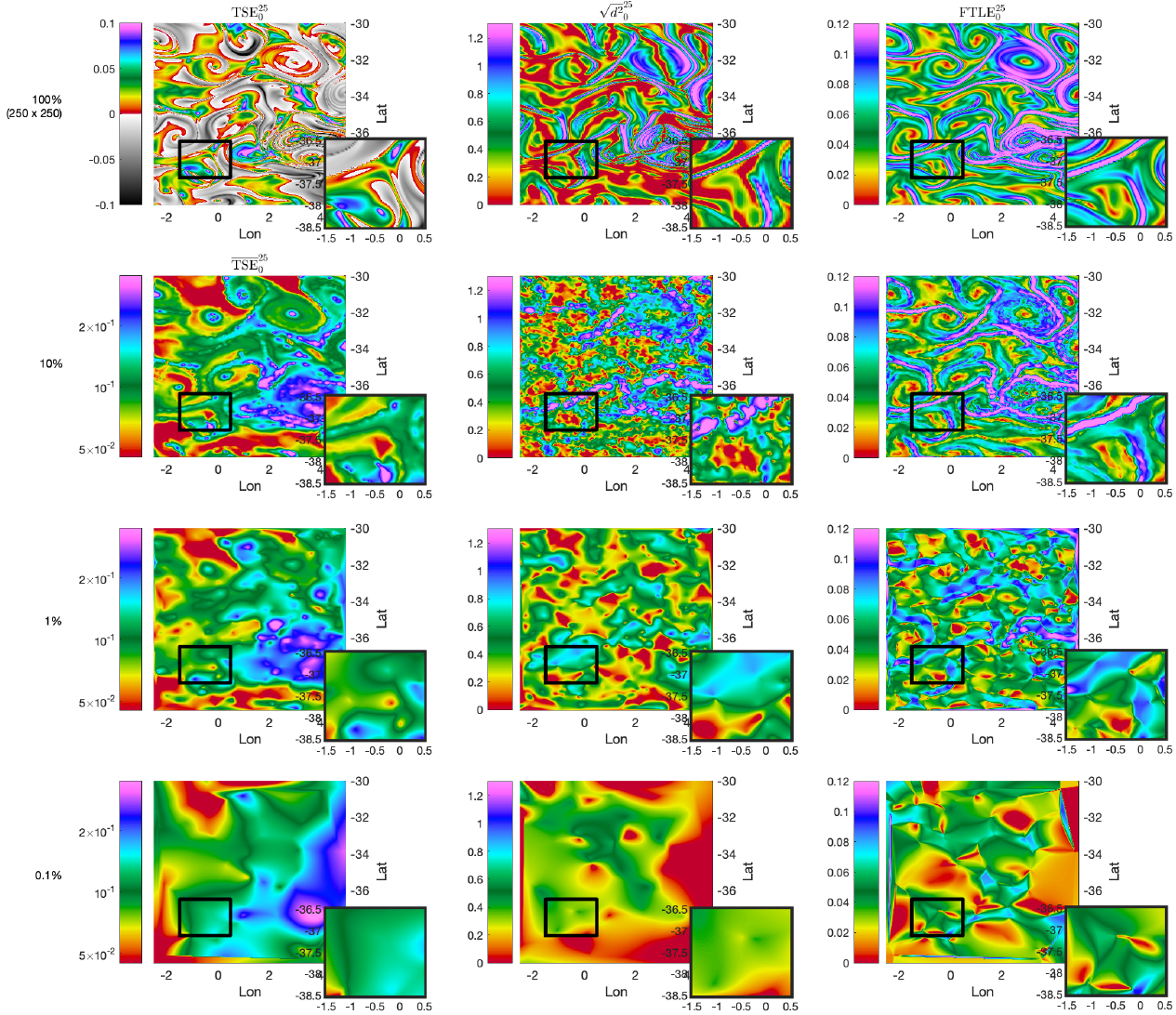}\caption{Stretching metric comparisons for AVISO ocean surface current fields.
The top row shows computations at full resolution with the lower rows
having progressively reduced resolution by randomly subsampling with
fewer trajectories. \label{fig:Stretching-metric-comparisons}}
\end{figure}

\begin{figure}[t]
\centering \includegraphics[width=1\textwidth]{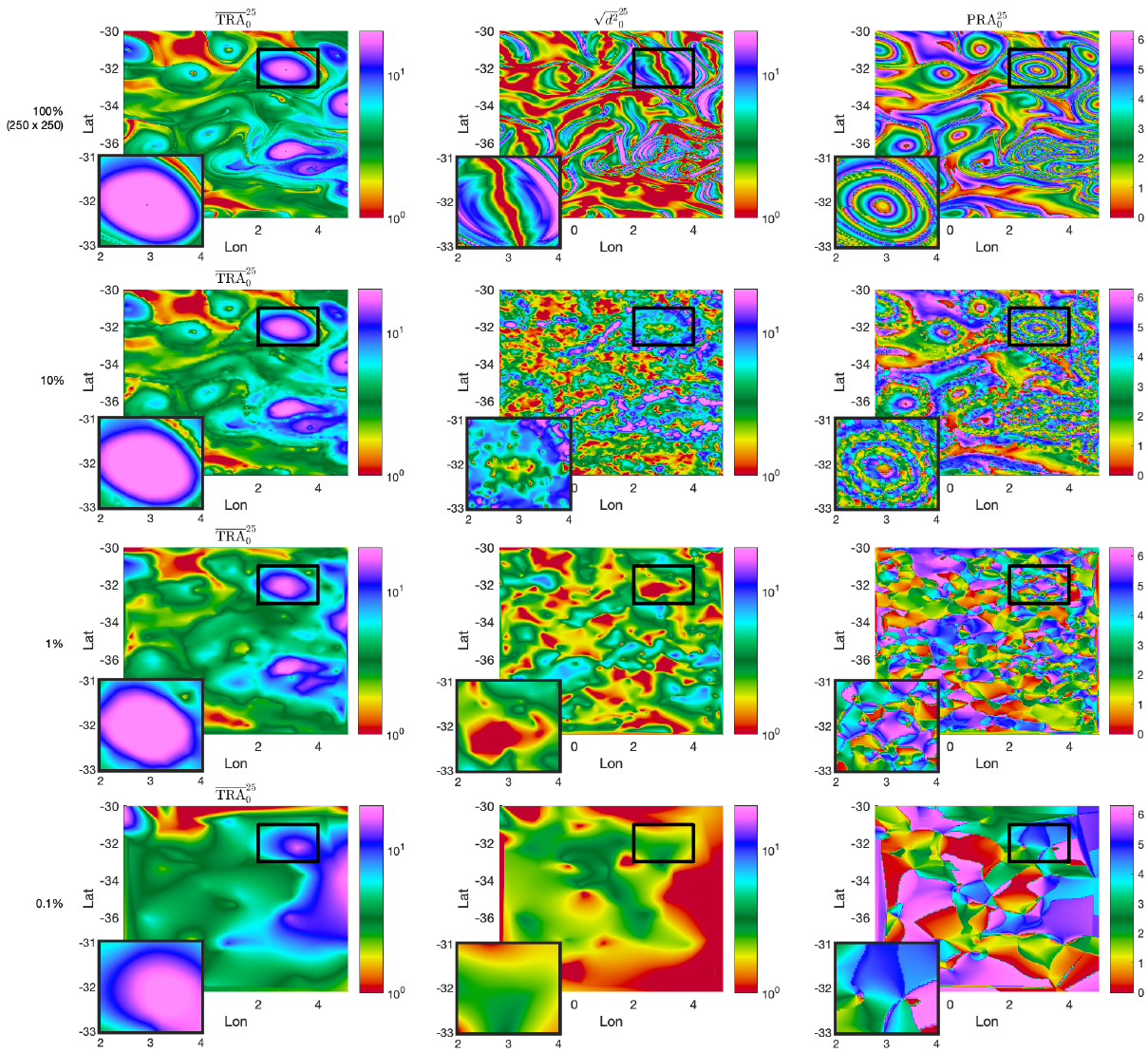}\caption{Rotation metric comparisons for AVISO ocean surface current fields.
The top row shows computations at full resolution with the lower rows
having progressively reduced resolution by randomly subsampling with
fewer trajectories. We display $\sqrt{d^{2}}$ here for clarity.\label{fig:Rotation-metric-comparisons}}
\end{figure}

\begin{figure}[t]
\centering \includegraphics[width=1\textwidth]{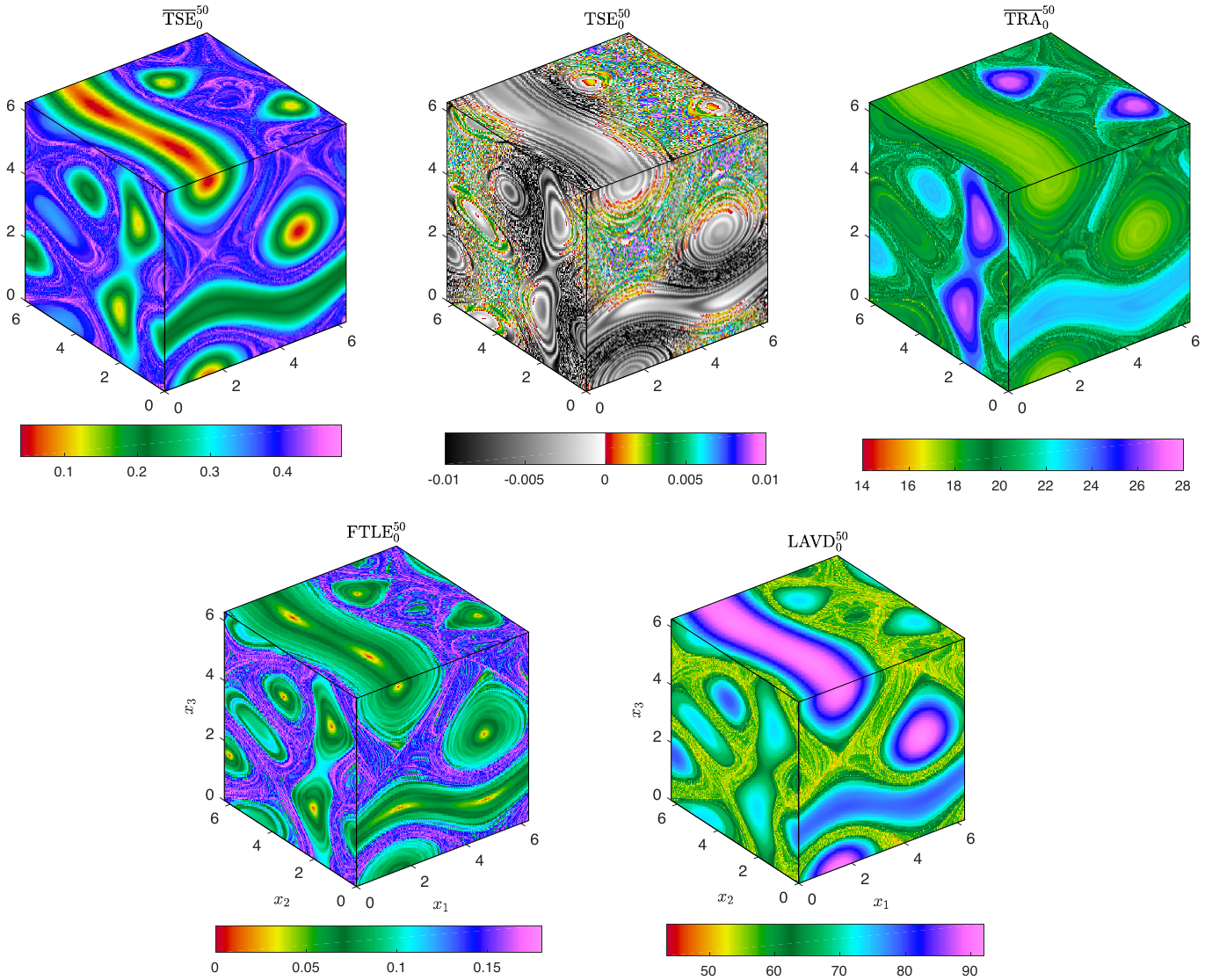}\caption{Elliptic and hyperbolic LCSs in the unsteady ABC flow (41). The plots
compare quasi-objective, single-trajectory metrics (\textbf{$\mathrm{TSE_{0}^{50}(\boldsymbol{x}\mathbf{}_{0})}$
}and\textbf{ $\mathrm{\mathrm{\overline{\mathrm{TRA}}{}_{0}^{50}}}(\boldsymbol{x}\mathbf{}_{0})$})
with objective LCS metrics (\textbf{$\mathrm{FTLE}_{0}^{50}(\boldsymbol{x}\mathbf{}_{0})$
}and\textbf{ $\mathrm{LAVD}_{0}^{50}(\boldsymbol{x}\mathbf{}_{0})$})
that require multiple neighboring trajectories or detailed knowledge
of the velocity field. \label{fig:ABC flow high res}}
\end{figure}

\begin{figure}[t]
\centering \includegraphics[width=1\textwidth]{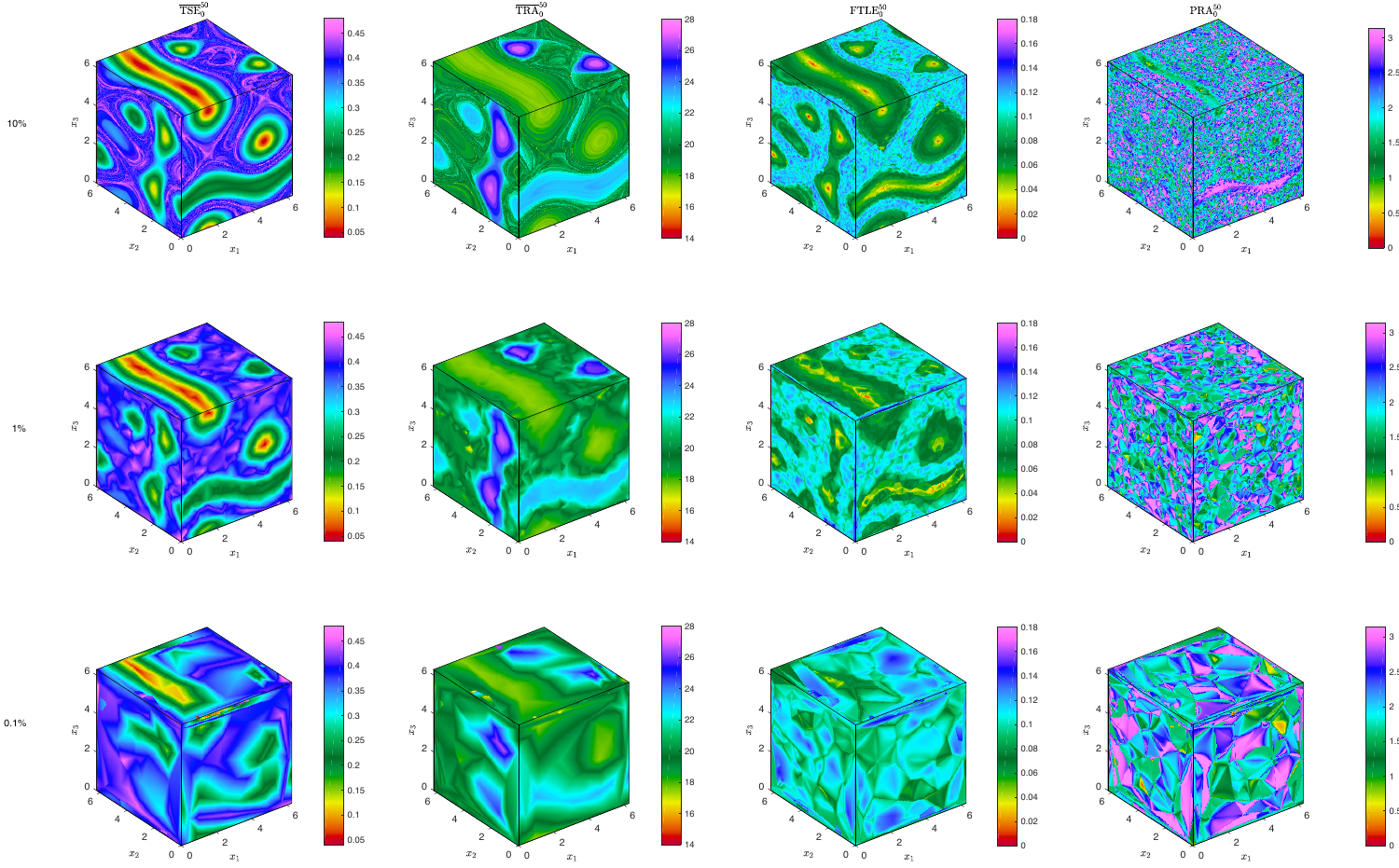}\caption{Elliptic and hyperbolic LCS in the randomly subsampled unsteady ABC
flow (41). The plots compare quasi-objective single-trajectory metrics
(\textbf{$\mathrm{\overline{TSE}{}_{0}^{50}}(\boldsymbol{x}\mathbf{}_{0})$
}and\textbf{ $\mathrm{\overline{TRA}_{0}^{50}}(\boldsymbol{x}\mathbf{}_{0})$})
with LCS metrics (\textbf{$\mathrm{FTLE}_{0}^{50}(\boldsymbol{x}\mathbf{}_{0})$
}and\textbf{ $\mathrm{PRA}_{0}^{50}(\boldsymbol{x}\mathbf{}_{0})$}),
whose computation requires multiple neighboring trajectories. The
initial condition grid for trajectories is randomized and its density
is gradually decreased to $0.1\%$ of its initial value. \label{fig:ABC flow sparsified}}
\end{figure}

\begin{singlespace}
We would like to thank Holger Theisel, Anke Friederici and Tobias
Günther for pointing out the invalidity of our original Theorem 3
via a counterexample in \cite{theisel22}. We are also grateful to
Bálint Kaszás for his helpful comments and for bringing Ref. \cite{pedlosky}
to our attention.
\end{singlespace}